\documentclass[twocolumn,showpacs,aps,prb,preprintnumbers,amsmath,amssymb]{revtex4}
\usepackage{graphicx}% Include figure files
\usepackage{dcolumn}% Align table columns on decimal point
\usepackage{bm}% bold math
\usepackage{epsfig}

\begin{document}
\date{\today}

%opening
\title{
Weak antilocalization in HgTe quantum wells and topological surface states: \\ Massive versus massless Dirac fermions    
}
\author{G. Tkachov and E. M. Hankiewicz}
\affiliation{
Institute for Theoretical Physics and Astrophysics,
W\"urzburg University, Am Hubland
97074 W\"urzburg, Germany}

%\affiliation{}

\begin{abstract}
HgTe quantum wells and surfaces of three-dimensional topological insulators support Dirac fermions with a single-valley band dispersion. 
In the presence of disorder they experience weak antilocalization, which has been observed in recent transport experiments. 
In this work we conduct a comparative theoretical study of the weak antilocalization in HgTe quantum wells and topological surface states. 
The difference between these two single-valley systems comes from a finite band gap (effective Dirac mass) in HgTe quantum wells 
in contrast to gapless (massless) surface states in topological insulators. The finite effective Dirac mass implies a broken internal symmetry, 
leading to suppression of the weak antilocalization in HgTe quantum wells at times larger than certain $\tau_{\cal M}$, inversely proportional to the Dirac mass.  
This corresponds to the opening of a relaxation gap $\tau^{-1}_{\cal M}$ in the Cooperon diffusion mode which we obtain from the Bethe-Salpeter equation 
including relevant spin degrees of freedom.  We demonstrate that the relaxation gap exhibits 
an interesting nonmonotonic dependence on both carrier density and band gap, vanishing at a certain combination of these parameters. 
The weak-antilocalization conductivity reflects this nonmonotonic behavior which is unique to HgTe QWs and absent for topological surface states. 
On the other hand, the topological surface states exhibit specific weak-antilocalization magnetoconductivity 
in a parallel magnetic field due to their exponential decay in the bulk.    
\end{abstract}
\maketitle

\section{Introduction}
\label{intro}

Recently discovered materials -- graphene,~\cite{Novoselov04,Neto09} 
two-dimensional (2D)~\cite{Kane05,Bernevig06,Koenig07,Roth09} and three-dimensional (3D)~\cite{Fu07,Moore07,Hsieh08,Hsieh09,Xia09,Chen09} 
topological insulators (TIs)~\cite{Hasan10,Qi10} -- exhibit a Dirac-like band dispersion which is responsible for their unusual electronic and optical properties. 
In graphene the low-energy electron spectrum can be approximated by two spin-degenerate Dirac cones at the corners of the Brillouin zone.~\cite{Semenov84} 
The 2D TIs have been realized in HgTe quantum wells (QWs)~\cite{Bernevig06,Koenig07} 
which have a single double-degenerate Dirac valley, as predicted by band-structure calculations and inferred from transport measurements.~\cite{Buettner11} 
The double degeneracy of the HgTe QW bands allows for an energy gap at the Dirac point, 
without breaking time-reversal invariance, which paves the way to study Dirac fermions with a finite (positive and negative) 
effective mass and related mass disorder.~\cite{GT11,GT10a,GT10b} In comparison with HgTe QWs, the ideal 
3D TI exhibits a single {\em non-degenerate} gapless Dirac cone on the surface of the material, 
whereas its bulk is insulating.~\cite{Volkov85} 
In this case, the opening of the gap in the surface Dirac spectrum requires time-reversal symmetry breaking and 
has been predicted to cause the surface quantum Hall effect~\cite{Qi08,Essin09,Tse10a,Tse10b} and 
rich magneto-electric phenomena~\cite{Qi08,Essin09,Tse10a,Tse10b,Maciejko10,Garate10,GT10c} related to axion electrodynamics.~\cite{Wilczek87}

The number of the Dirac valleys is an essential factor in understanding quantum electron transport in disordered samples. 
The transport studies of graphene have reported weak localization~\cite{Morozov06} and more complex quantum-interference patterns~\cite{Tikhonenko09} 
instead of the antilocalization effect expected for the symplectic universality class 
(e.g. Refs.~\onlinecite{Hikami80,Iordanskii94,Lyanda-Geller94,Knap96,Altland97,Aleiner01,Brouwer02,Suzuura02,Miller03,Zaitsev05,McCann06,Aleiner06,Kechedzhi07,Neumaier07,Guzenko07,Imura09,Wurm09}). 
Such a situation can occur if the two graphene's valleys are coupled by scattering off atomically sharp defects.~\cite{Suzuura02,McCann06,Kechedzhi07,Imura09,Wurm09} 
In contrast, in single-valley Dirac systems such scattering processes are forbidden, and recent experiments on HgTe QWs~\cite{Olshanetsky10,Buettner11WAL} 
and 3D TIs~\cite{Checkelsky10, Chen10, He10} have found a positive (antilocalization) quantum-interference conductivity. 

In this work we conduct a comparative study of the weak antilocalization (WAL) in HgTe QWs and on surfaces of 3D TIs. 
The goal is to elucidate the difference between these two systems which comes from the finite effective Dirac mass in HgTe QWs 
in contrast to the massless surface states in 3D TIs. Like in the conventional 2D electron systems (2DESs) 
with Bychkov-Rashba or Dresselhaus spin-orbit interactions,~\cite{Iordanskii94,Knap96} 
the WAL in HgTe QWs and on surfaces of 3D TIs reflects the broken rotation symmetry in relevant spin space. 
However, in addition to the lack of this continuous symmetry, 
the effective Dirac mass in HgTe QWs implies a broken discrete symmetry which for a single-cone system would play the role of time reversal. 
In this sense, there is a formal analogy between the effective Dirac mass and an out-of-plane Zeeman field in a 2DES.~\cite{Neumaier07,Dugaev01} 
Therefore, by analogy with weak ferromagnets~\cite{Neumaier07,Dugaev01} we find that the WAL in HgTe QWs is suppressed at times larger than certain $\tau_{\cal M}$, 
inversely proportional to the effective Dirac mass. Such suppression is however absent for the massless surface states in 3D TIs, 
which can be used to experimentally differentiate between the two systems. 

Before going to the calculation details given in Sec.~\ref{s_HgTe} and \ref{s_BiSe}, in the next section  
we briefly announce some of our results for HgTe QWs and TIs.  

\section{Overview of the results}
\label{over}

\begin{figure}[t]
\includegraphics[width=70mm]{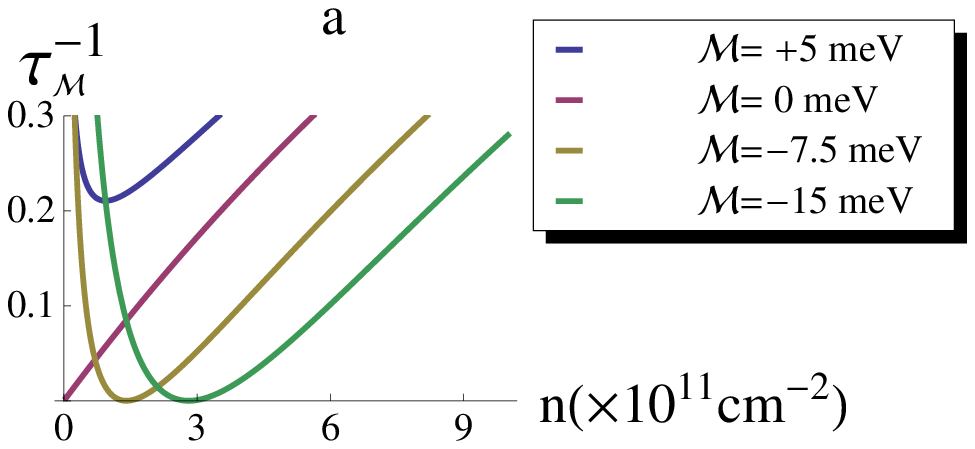}
\includegraphics[width=70mm]{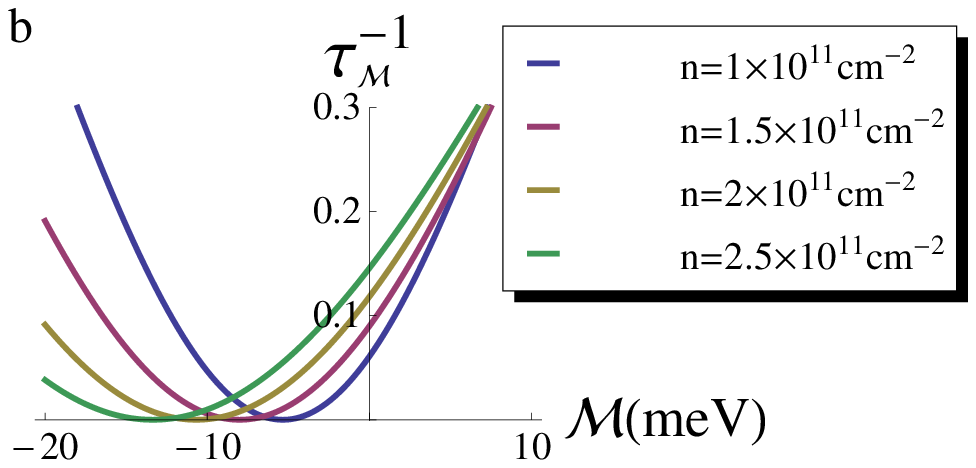}
\caption{ (Color online)
Relaxation gap $\tau^{-1}_{\cal M}$ [in units of inverse life-time $\tau^{-1}$, see Eq.~(\ref{Gamma})] versus carrier density $n$ (a) 
and band gap ${\cal M}$ (b); ${\cal A}=380$ meV$\cdot$nm and ${\cal B}=850$ meV$\cdot$nm$^2$ (from Ref.~\onlinecite{Buettner11}). 
}
\label{Gap}
\end{figure}

\begin{figure}[b]
\includegraphics[width=75mm]{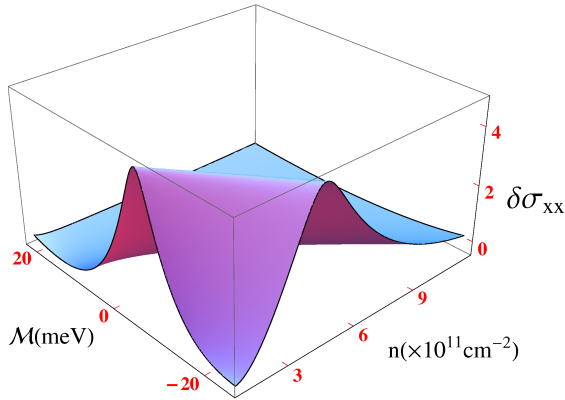}
\caption{(Color online)
Quantum-interference conductivity correction $\delta\sigma_{xx}$ [in units of $e^2/\pi h$, see Eq.~(\ref{dS_HgTe})] 
versus carrier density $n$ and band gap ${\cal M}$; ${\cal A}=380$ meV$\cdot$nm, ${\cal B}=850$ meV$\cdot$nm$^2$ 
(from Ref.~\onlinecite{Buettner11}) and $\tau_0/\tau_\varphi = 0.01$ [see also Eq.~(\ref{tau_HgTe}) for $\tau$ and $\tau_0$ and Eq. (\ref{alpha}) for $\alpha$]. 
}
\label{dS_nM}
\end{figure}

To calculate the quantum-interference (Cooperon) conductivity correction, $\delta\sigma_{xx}$, we adopt the effective Dirac model of HgTe QWs~\cite{Bernevig06} and obtain the following expression for $\delta\sigma_{xx}$:        
\begin{eqnarray}
&&
\delta\sigma_{xx}(n,{\cal M}) = - \alpha \frac{2e^2} {\pi h} \ln\frac{\tau^{-1}}{ \tau^{-1}_{\cal M} + \tau^{-1}_\varphi },
\label{dS_HgTe}\\
%\end{eqnarray}
%
%\begin{eqnarray} 
&&
\tau^{-1}_{\cal M} = 
\frac{2}{\tau} \frac{ ( {\cal M} + {\cal B}k^2_{_F}  )^2  }{ {\cal A}^2k^2_{_F} + ( {\cal M} + {\cal B}k^2_{_F} )^2 }, 
\quad 
k_{_F}=\sqrt{2\pi n}.
\label{Gamma}
\end{eqnarray}
Here the symmetry-breaking-induced relaxation gap $\tau^{-1}_{\cal M}$ is 
proportional to the total mass term, ${\cal M} + {\cal B}k^2_{_F}$, in the effective Dirac model of HgTe QWs.~\cite{Bernevig06} ${\cal M}$ 
is the band gap at the Dirac point,  ${\cal B}k^2_{_F}$ is the positive quadratic correction accounting for the curvature of the filled conduction band 
($k_{_F}$ is the Fermi momentum determined by the 2D carrier concentration $n$), whereas constant ${\cal A}$ determines the linear (Dirac) part of the spectrum 
($\tau$ the elastic life-time). 
In Eq.~(\ref{dS_HgTe}) the factor of 2 accounts for the double degeneracy of the Dirac valley in HgTe QWs,
constant $\alpha$ approaches $-1/2$ for ${\cal M} + {\cal B}k^2_{_F}\to 0$ (as discussed in detail in Sec. \ref{s_HgTe}) and $\tau_{\varphi}$ is the dephasing time.  

\begin{figure}[t]
\includegraphics[width=65mm]{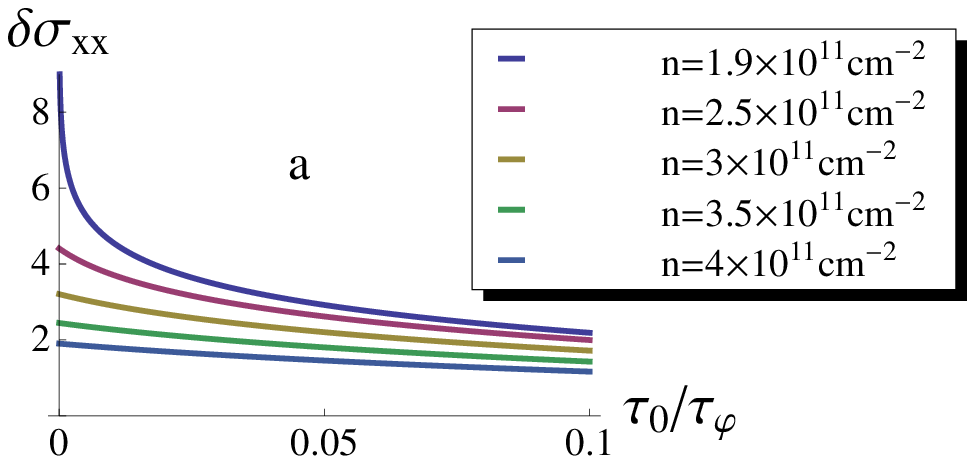}
\includegraphics[width=65mm]{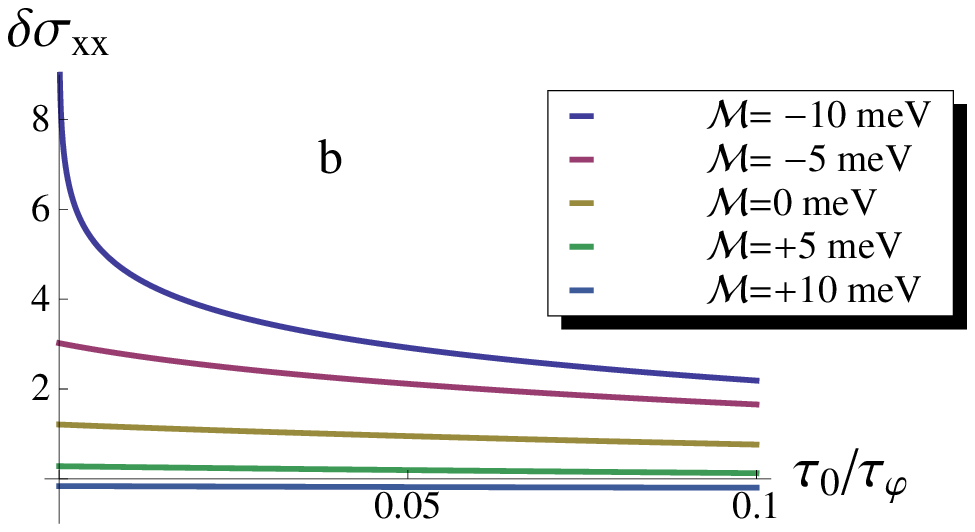}
\caption{ (Color online)
Quantum-interference conductivity correction $\delta\sigma_{xx}$ [in units of $e^2/\pi h$, see Eq.~(\ref{dS_HgTe})] 
versus normalized dephasing rate $\tau_0/\tau_\varphi$ for different carrier densities and ${\cal M}=-10$ meV (a) 
and for different band gaps and $n=1.9 \times 10^{11}$ cm$^{-2}$ (b); ${\cal A}=380$ meV$\cdot$nm and ${\cal B}=850$ meV$\cdot$nm$^2$ (from Ref.~\onlinecite{Buettner11}). See also Eq.~(\ref{tau_HgTe}) for $\tau$ and $\tau_0$ and Eq. (\ref{alpha}) for $\alpha$.  
}
\label{dS_deph}
\end{figure}

The unique feature of the HgTe QWs is that the band gap ${\cal M}$ can take both positive and negative values depending on the QW thickness.~\cite{Koenig07,Buettner11} 
Therefore, the relaxation gap $\tau^{-1}_{\cal M}$ (\ref{Gamma}) exhibits an interesting nonmonotonic behavior as a function of both ${\cal M}$ 
and carrier density $n$ [see  Fig.~\ref{Gap}], vanishing when these parameters satisfy the condition, 
\begin{equation}
 {\cal M} + 2\pi n {\cal B} = 0.
 \label{maximum}
\end{equation}
It represents a line in parameter space $({\cal M}, n)$ on which conductivity (\ref{dS_HgTe}) 
reaches the maximum $\delta\sigma_{xx}=(e^2/\pi h)\ln (\tau_\varphi/\tau )$ 
[see also Fig.~\ref{dS_nM}]. Such a nonmonotonic behavior of $\delta\sigma_{xx}(n,{\cal M})$ 
can be used to experimentally identify the symmetry breaking and the resulting relaxation gap $\tau^{-1}_{\cal M}$. 
In particular, the predicted carrier-density dependence should hold for the QWs where the potential impurity scattering is much stronger than 
scattering off random gap fluctuations. This regime can be identified from the carrier-density dependence of the QW mobility.~\cite{GT11}  
As to the dependence on the gap $M$, it can be extracted from sample-to-sample measurements.  
The band structure of MBE grown HgTe QWs is controllable to a great extent.~\cite{Koenig07,Roth09,Buettner11}  
For the experiment we suggest here one should select several QWs with distinctly different gaps 
and comparable dephasing times (inferred from the temperature dependence of the conductivity) and 
other band structure parameters (inferred from the Hall and Shubnikov-de Haas measurements). 
Alternatively, the presence of the relaxation gap $\tau^{-1}_{\cal M}$ can be established from the dependence of $\delta\sigma_{xx}$ 
on the dephasing rate $1/\tau_{\varphi}$, which is directly related to the temperature dependence (e.g.  
the dephasing rate due to electron-electron interactions is linearly proportional to the temperature).~\cite{Aleiner99}   
The dependence of $\delta\sigma_{xx}$ on $1/\tau_{\varphi}$ is shown in Figs.~\ref{dS_deph}(a) and (b).
In these figures the upper curves correspond to the logarithmically divergent $\delta\sigma_{xx}$ with $\tau^{-1}_{\cal M}=0$. 
In contrast, for finite $\tau^{-1}_{\cal M}$ (rest of the curves) the conductivity correction shows only weak dependence on the dephasing rate.

As to the 3D TIs, we focus on compounds Bi$_2$Se$_3$ and Bi$_2$Te$_3$ where the surface states can be described by the effective two-band Dirac Hamiltonian, 
accounting for the hexagonal warping of the bands [see, e.g., Ref.~\onlinecite{Fu09,Liu10}]. 
The warping term is cubic in momentum ${\bf k}$ and enters formally as the Dirac mass term. 
However, since it preserves the time-reversal symmetry, we find that for weak warping the quantum-interference conductivity correction 
has the same form as for the conventional 2DES with spin-orbit interaction:~\cite{Hikami80,Iordanskii94,Knap96} 
\begin{eqnarray}
\delta\sigma_{xx}=-\alpha\frac{e^2}{\pi h}\ln\frac{ \tau_\varphi }{ \tau }, \qquad \alpha=-\frac{1}{2}.
\label{dS_BiSe}
\end{eqnarray}
The specific of the surface state shows up mainly in the dependence of the conductivity 
$\Delta\sigma_{xx}(B)=\delta\sigma_{xx}(B)-\delta\sigma_{xx}(0)$ on magnetic field $B$ applied {\em parallel} to the TI surface:
\begin{eqnarray}
\Delta\sigma_{xx}(B) = -\frac{e^2}{2\pi h}\ln\Biggl( 1 + \frac{B^2}{ B^2_{_\|} } \Biggr),\quad
B_{_\|} = \frac{\hbar}{e\lambda \sqrt{\ell_{tr} \ell_\varphi}}.
\label{dsigma_in1}
\end{eqnarray}
Such dependence reflects the finite penetration length, $\lambda$, of the surface state into the bulk, 
i.e. the magnetic flux through the effective width of the surface state [see Eq. (\ref{dsigma_in1}) for $B_{_\|}$, 
where $\ell_{tr}$ and $\ell_\varphi$ are the transport mean free path and dephasing length, respectively]. 
Quantum transport in the in-plane magnetic fields has been studied theoretically for disordered metallic films~\cite{Altshuler81} 
and electron quantum wells.~\cite{Meyer02} The present case differs from the previous studies in that the topological surface states  
have a different microscopic profile of the transverse wave functions.
We discuss the dependence of $\Delta\sigma_{xx}$ on the magnetic field orientation in Sec.~\ref{s_BiSe} 
in connection with recent experiments on Bi$_2$Te$_3$ (Ref. \onlinecite{He10}).

\begin{figure}[t!]
\includegraphics[width=40mm]{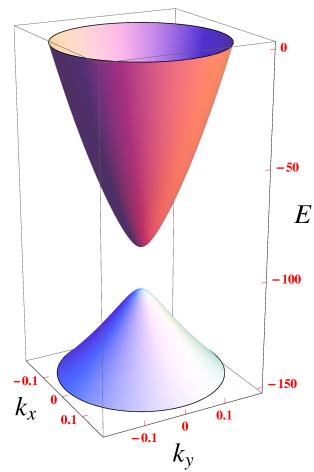}
\caption{(Color online)
Energy bands of a HgTe quantum well [see Eq. (\ref{H}) and text] in meV versus in-plane wave-numbers $k_x$ and $k_y$ in nm$^{-1}$. 
The Fermi level lies in the conduction band at $E=0$. We choose ${\cal A}=380$ meV$\cdot$nm, ${\cal B}=850$ meV$\cdot$nm$^2$, ${\cal D}=700$ meV$\cdot$nm$^2$ 
(from Ref.~\onlinecite{Buettner11}), and $E_{_F}=100$ meV.
}
\label{E}
\end{figure}

\section{HgTe quantum wells}
\label{s_HgTe}

\subsection{Effective Hamiltonian}
\label{ss_H_HgTe}

We use the effective 4-band Hamiltonian of HgTe wells~\cite{Bernevig06} 
which can be written in the following form:   

\begin{eqnarray}
H_{HgTe}=
\left(
\begin{array}{cc}
H + H_i & 0\\
0 & \tilde{H} + H_i
\end{array}
\right),
\label{H_HgTe}
\end{eqnarray}
\begin{eqnarray}
&
H=
\mbox{\boldmath$\sigma$}( {\cal A} {\bf\hat k} + {\cal M}_{\bf\hat k}{\bf z} ) 
+  ( {\cal D}{\bf\hat k}^2 - E_{_F} )\sigma^0, 
&
\label{H}\\
&
\tilde{H}=
-\mbox{\boldmath$\sigma$}( {\cal A} {\bf\hat k} + {\cal M}_{\bf\hat k}{\bf z} ) 
+  ( {\cal D}{\bf\hat k}^2 - E_{_F} )\sigma^0, 
&
\label{H_tilde}\\
&
H_i=V({\bf r}) \sigma^0.
&
\label{H_i}
\end{eqnarray}
The two diagonal blocks in $H_{HgTe}$ describe pairs of states related to each other by time reversal symmetry (Kramers partners). 
In the upper block the Hamiltonian $H$ has a matrix $2 \times 2$ structure 
with Pauli matrices $\sigma^{x,y,z}$ and unit matrix $\sigma^0$ acting in space of two lowest-energy subbands of 
the HgTe quantum well:~\cite{Bernevig06} 
an s-like electron one $|E_1, \frac{1}{2} \rangle$ and a p-like heavy hole one $|H_1, \frac{3}{2} \rangle$. 
For the lower block the basis states have the opposite spin projections: 
$|E_1, -\frac{1}{2} \rangle$ and $|H_1, -\frac{3}{2} \rangle$.
The linear term in $H$ (proportional to constant ${\cal A}$ and momentum operator ${\bf\hat k}=-i\nabla_{\bf r}$) 
originates from the hybridization of the s- and p-like subbands.    
${\cal M}_{\bf\hat k}$ is the effective Dirac mass: 
\begin{eqnarray}
{\cal M}_{\bf\hat k }={\cal M} + {\cal B}{\bf\hat k }^2, 
\label{M_HgTe}
\end{eqnarray}
where constant ${\cal M}$ determines the band gap at the gamma (${\bf k}=0$) point of the Brillouin zone (see Fig. \ref{E}).  
The positive quadratic terms ${\cal B}{\bf\hat k}^2$ and ${\cal D}{\bf\hat k}^2$ take into account 
the details of the band curvature in HgTe quantum wells.~\cite{Bernevig06,Koenig07} 
Finally, $H_i$ in Eq.~(\ref{H_HgTe}) accounts for interaction with randomly distributed short-range impurities. 
The impuritity potential $V({\bf r})$ is characterized by the correlation function, 
\begin{eqnarray}
&&
\zeta({\bf r} - {\bf r}^\prime)=\langle\langle V({\bf r})V({\bf r}^\prime) \rangle\rangle 
=\frac{\hbar}{\pi N \tau_0}\, \delta( {\bf r} - {\bf r}^\prime ),
\label{Corr}\\
&&
\zeta_{\bf k}=\int \zeta({\bf r}) {\rm e}^{-i{\bf k}{\bf r} } d{\bf r} = \frac{\hbar}{\pi N \tau_0},
\label{Corr_k}
\end{eqnarray}
parametrized in terms of the characteristic scattering time $\tau_0$ and 
the density of states (DOS) at the Fermi level, $N$, for one Dirac cone 
[brackets $\langle\langle...\rangle\rangle$ denote averaging over impurity positions and 
$\zeta_{\bf k}$ is the Fourier transform of the disorder correlation function].

We emphasize that the mass term ${\cal M}_{\bf\hat k}$ (\ref{M_HgTe}) is symmetric 
with respect to momentum inversion ${\bf\hat k} \to - {\bf\hat k}$. 
Hence, Hamiltonian $H$ does not possess the symmetry under transformation 
\begin{equation}
{\bf\hat k},\mbox{\boldmath$\sigma$} \to -{\bf\hat k},-\mbox{\boldmath$\sigma$}.
\label{Trans} 
\end{equation}
Within a given block of Eq.~(\ref{H_HgTe}), i.e. in subband pseudospin space, 
such a transformation plays the role of "time reversal". 
At the same time, the real time-reversal symmetry is ensured by the matrix form 
of $H_{HgTe}$ (\ref{H_HgTe}). Physically, this means that the Kramers partners 
reside on two Dirac cones superimposed at ${\bf k}=0$ point.~\cite{Buettner11}
Note that the zero off-diagonal elements in $H_{HgTe}$ (\ref{H_HgTe}) imply 
conservation of the spin projections of $|E_1 \rangle$ and $|H_1 \rangle$ subbands, 
which is a good approximation for symmetric HgTe wells.~\cite{Koenig07,Buettner11,Rothe10,BIA}
In this case, each of the Dirac cones contributes independently to transport processes, 
which is accounted for by the factor of 2 in the expressions for the conductivity corrections 
discussed in subsection~\ref{ss_ds_HgTe}.

\subsection{Disorder-averaged single-particle Green's functions and elastic life-time}
\label{ss_G_HgTe}

\begin{figure}[t!]
\includegraphics[width=85mm]{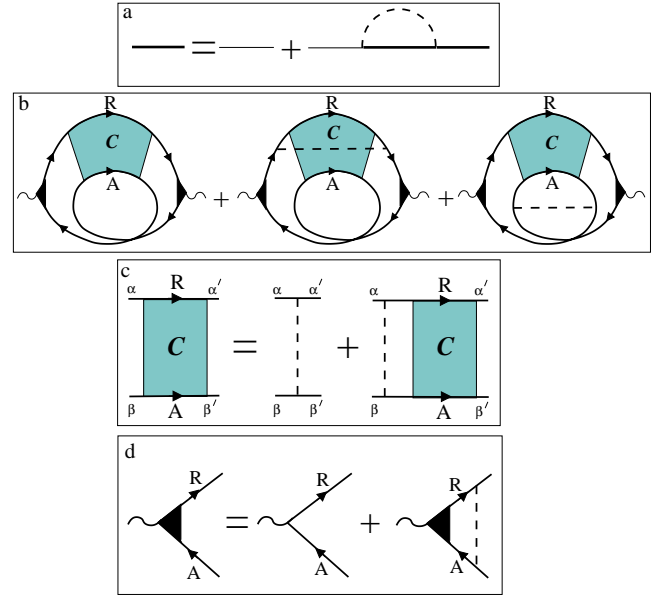}
\caption{(Color online)
Diagrammatic representations for
(a) equation for disorder-averaged Green's function (thick line) in 
self-consistent Born approximation (thin line is the unperturbed Green's function of the disorder-free system, 
dashed line is the disorder correlator), 
(b) bare and dressed Hikami boxes for the Cooperon correction to Drude conductivity, 
(c) Bethe-Salpeter equation for the Cooperon, and 
(d) equation for the renormalized current vertex in the ladder approximation. 
}
\label{Diagrams}
\end{figure}

We begin by calculating the disorder-averaged retarded/advanced Green's functions $G^{^{R/A}}$ for 
an n-type HgTe well under weak-scattering condition
\begin{equation}
 k_{_F}v_{_F}\tau \gg 1,
\label{weak}
\end{equation}
where $\tau$ is the elastic scattering time and $v_{_F}$ and $k_{_F}=\sqrt{2\pi n}$
are the Fermi velocity and wave-vector, respectively.
In the standard self-consistent Born approximation (see, e.g. Refs. \onlinecite{AGD,Rammer}) 
the equation for $G^{^{R/A}}$ is shown diagrammatically in Fig.~\ref{Diagrams}(a). In ${\bf k}$ representation it reads 
\begin{eqnarray}
G^{^{R/A}}_{\bf k}=G^{^{R/A} }_{0 \bf k} +
G^{^{R/A} }_{0 \bf k} \int \frac{ d{\bf k^\prime} }{(2\pi)^2} 
\zeta_{\bf k-k^\prime}G^{^{R/A} }_{\bf k^\prime}\, 
G^{^{R/A}}_{\bf k},
\label{Dyson_G}
\end{eqnarray}
\begin{eqnarray}
G^{^{R/A}}_{0 \bf k}=\frac{1}{2}\frac{\sigma^0 + \mbox{\boldmath$\sigma$} {\bf e}_{\bf k} }
{\epsilon - \xi_{\bf k } }, 
{\bf e}_{\bf k}=
s\frac{  {\cal A} {\bf k} + {\cal M}_{\bf k } {\bf z} }{ \sqrt{  {\cal A}^2{\bf k}^2 +  {\cal M}^2_{\bf k }  }  },\, 
|{\bf e}_{\bf k}|^2=1,\,\,\,
\label{G_0} 
\end{eqnarray}
Here the Green's function $G^{^{R/A}}_{0 \bf k}$ describes a conduction-band carrier 
with dispersion $\xi_{\bf k }=\sqrt{  {\cal A}^2{\bf k }^2 +  {\cal M}^2_{\bf k }} + {\cal D}{\bf k }^2 - E_{_F}$ 
in the absence of disorder [index $s=\pm 1$ labels the Kramers partners residing on the different Dirac cones]. 
The valence band contribution is neglected under assumption that the energy separation between the valence and conduction bands 
is much bigger than the disorder-induced band smearing,
\begin{eqnarray}
2\sqrt{ {\cal A}^2k^2_{_F} + {\cal M}^2_{ k_F\cdot{\bf n}} } = 2( E_{_F} - {\cal D}k^2_{_F} ) \gg \hbar/\tau.
 \label{v-band}
\end{eqnarray}
Because of the large band-structure constant ${\cal A}\approx 380$ meV$\cdot$nm [see, e.g. Ref. \onlinecite{Buettner11}]
the requirement (\ref{v-band}) can be satisfied simultaneously with the weak-scattering condition (\ref{weak}).
We also note that the matrix structure of $G^{^{R/A}}_{0 \bf k}$ (\ref{G_0}) accounts for the carrier chirality and is of primary importance throughout the paper.  

The solution to Eq.~(\ref{Dyson_G}) can be sought in the form
$G^{^{R/A}}_{\bf k}=\frac{1}{2}(\sigma^0 + \mbox{\boldmath$\sigma$} {\bf e}_{\bf k})g^{^{R/A}}_{\bf k}$, 
where $g^{^{R/A} }_{\bf k}=1/( \epsilon - \xi_{\bf k } -\Sigma^{^{R/A} }_{\bf k} )$ and    
$\Sigma^{^{R/A}}_{\bf k}$ satisfies the equation  
\begin{eqnarray}
\Sigma^{^{R/A}}_{\bf k} =\int 
\frac{ \zeta_{\bf k-k^\prime} }
{\epsilon - \xi_{\bf k^\prime } -\Sigma^{^{R/A}}_{\bf k^\prime} }
\frac{1 + {\bf e}_{\bf k}{\bf e}_{\bf k^\prime} }{2}  \frac{ d{\bf k^\prime} }{ (2\pi)^2 }.
\label{Sigma}
\end{eqnarray}
Approximate solution~\cite{AGD,Rammer} of the latter equation near the Fermi surface, 
$|{\bf k}| \approx k_{_F}$, yields the disorder-averaged Green's function as 
\begin{eqnarray}
G^{^{R/A} }_{\bf k}=\frac{1}{2}
\frac{\sigma^0 + \mbox{\boldmath$\sigma$}{\bf e}_{ k_F\cdot{\bf n} } }
{\epsilon_{_{R/A} } - \xi_{\bf k } },
\qquad
\epsilon_{_{R/A} }=\epsilon \pm \frac{ i\hbar}{2\tau},
\label{G_RA}
\end{eqnarray}
where $\tau$ is the elastic life-time given by
\begin{eqnarray}
&&
\frac{1}{\tau}=\mp \frac{2}{\hbar} {\rm Im}\,\Sigma^{^{R/A}}_{ k_F \cdot \bf{n} } =
\frac{N}{\hbar}\times
\label{tau}\\
&&
\times
\int d\phi_{ {\bf n}^\prime } 
\zeta_{ k_F \cdot |{\bf n} - {\bf n}^\prime | } 
\frac{
1  + {\bf e}_{\perp {\bf n}} \cdot {\bf e}_{\perp {\bf n}^\prime}
   + {\bf e}_{_\| {\bf n} } \cdot {\bf e}_{_\| {\bf n}^\prime} 
}{2},
\nonumber
\end{eqnarray}
\begin{eqnarray}
{\bf e}_{_\| {\bf n}} = s {\bf n} \sqrt{1 - {\bf e}^2_{\perp {\bf n}} },
\quad
{\bf e}_{\perp {\bf n}} = s \frac{ {\cal M}_{k_F\cdot{\bf n}}\, {\bf z} }
                                { \sqrt{ {\cal A}^2k^2_{_F} + {\cal M}^2_{ k_F\cdot{\bf n}} } }.
\label{e_pp}
\end{eqnarray}
In Eq.~(\ref{tau}) the unit vectors ${\bf n}$ and ${\bf n}^\prime$ specify the directions of the incident and scattered 
momentum states, respectively, and ${\bf e}_{_\| {\bf n}}$ and ${\bf e}_{\perp {\bf n}}$ are the in- and out-of-plane 
components of the unit vector ${\bf e}_{ k_F\cdot{\bf n} }={\bf e}_{_\| {\bf n}}+{\bf e}_{\perp {\bf n}}$.  
For isotropic $\zeta_{ k_F \cdot |{\bf n} - {\bf n}^\prime | }$ (\ref{Corr_k}) and 
${\cal M}_{k_F\cdot{\bf n}}$ (\ref{M_HgTe}) one finds the elastic time  
\begin{eqnarray}
\tau = \frac{\tau_0}{1 + {\bf e}^2_\perp}, 
\label{tau_HgTe}
\end{eqnarray}
which is shorter than the disorder-related time scale $\tau_0$ [see, Eq.~(\ref{Corr})]. 
This is due to the allowed backscattering into an opposite momentum state (${\bf n} {\bf n}^\prime \approx -1$) 
which is absent in the gapless case.~\cite{Ando05} 
The backscattering is the consequence of the symmetry breaking due to the mass term. 
The strength of the symmetry breaking is controlled by the out-of-plane component ${\bf e}_{\perp {\bf n}}$ 
of the unit vector ${\bf e}_{ k_F\cdot{\bf n} }$ [see, Eq.~(\ref{e_pp})].

\subsection{Cooperon}
\label{ss_C_HgTe}

The quantum-interference corrections to the Drude conductivity can be expressed diagrammatically by the Hikami boxes 
shown in Fig~\ref{Diagrams}b. Apart from the single-particle Green's functions $G^{^{R/A} }_{\bf k}$ (\ref{G_RA})
they involve the disorder-averaged two-particle Green's function, $C_{\alpha\beta\alpha^\prime\beta^\prime}({\bf q})$,
known as the Cooperon. In this subsection we will set up and solve the equation for $C_{\alpha\beta\alpha^\prime\beta^\prime}({\bf q})$. 
 
For the potential disorder defined by Eq.~(\ref{Corr}) 
the diagram for the Cooperon equation (see, Fig~\ref{Diagrams}c) is read off as follows
\begin{eqnarray}
&&
C_{\alpha\beta\alpha^\prime\beta^\prime}({\bf q})= 
\frac{\tau^2}{\tau_0}\delta_{\alpha\alpha^\prime}\delta_{\beta\beta^\prime} 
+ 
\frac{\hbar}{\pi N \tau_0}
\int \frac{d{\bf k}}{(2\pi)^2}
\nonumber\\
&&
\times
\sum_{\gamma^\prime\delta^\prime}
G^{^R}_{\alpha\gamma^\prime}({\bf k}, \epsilon +\hbar\omega )
G^{^A}_{\beta\delta^\prime}({\bf q - k}, \epsilon) 
C_{\gamma^\prime\delta^\prime\alpha^\prime\beta^\prime}({\bf q}),\quad
\label{Eq}
\end{eqnarray}
where the Greek indices label the states in pseudospin ($\mbox{\boldmath$\sigma$}$) space. 
The prefactor $\tau^2/\tau_0$ in the first term is due to the chosen normalization of $C_{\alpha\beta\alpha^\prime\beta^\prime}({\bf q})$.  
To solve Eq.~(\ref{Eq}) it is convenient to first expand the Cooperon 
in the orthonormal eigenfunctions of the pseudospin of a two-electron system:
\begin{eqnarray}
C_{\alpha\beta\alpha^\prime\beta^\prime}=\sum_{ij}
C^{ij}\, \Psi^i_{\alpha\beta}\Psi^{j*}_{\alpha^\prime\beta^\prime},
\,\,
\sum_{\alpha\beta}
\Psi^j_{\alpha\beta}\Psi^{i*}_{\alpha\beta}=\delta_{ji}. 
\label{Expansion}
\end{eqnarray}
The indices $i,j=0,x,y,z$ label the pseudospin-singlet ($0$) and pseudospin-triplet ($x,y,z$) states. 
The conductivity corrections will be eventually expressed in terms of the coefficients $C^{ij}$ 
for which we derive the following algebraic equations (see, also, Appendix~\ref{A}): 
\begin{eqnarray}
&&
C^{ij}({\bf q})= 
\frac{\tau^2}{\tau_0}\delta_{ij} 
+ 
\label{Eq4_C}\\
&& 
\frac{\tau}{4\tau_0} 
\sum_s
{\rm Tr}
\langle
(\sigma^0 - \mbox{\boldmath$\sigma$}{\bf e}_{-k_F\cdot{\bf n}} ) 
\sigma^i 
(\sigma^0 + \mbox{\boldmath$\sigma$}{\bf e}_{ k_F\cdot{\bf n}} ) 
\sigma^s
\rangle
C^{sj}({\bf q}),
\nonumber
\end{eqnarray}
where the square brackets stand for integral over the directions of the momentum ${\bf k}=k_{_F}{\bf n}$ 
on the Fermi surface: 
\begin{eqnarray}
&&
\langle...\rangle
=\int_0^{2\pi} \frac{ d\phi_{\bf n} }{2\pi}
\frac{...}
{1 - i \tau\omega + i\tau v_{_F}{\bf n}\cdot{\bf q} }. 
\label{Angle}
\end{eqnarray}
Evaluating the traces of the products of the Pauli matrices in Eq.~(\ref{Eq4_C}) we find  
\begin{eqnarray}
&&
C^{0j} = 
\frac{\tau^2}{\tau_0}\delta_{0j} 
+ 
\frac{\tau}{2\tau_0} 
\langle 
1 - {\bf e}_+\cdot {\bf e}_-    
\rangle
C^{0j}
\nonumber\\
&&
+ 
\frac{\tau}{2\tau_0} 
\sum\limits_{b=x,y,z}
\langle 
({\bf e}_+ - {\bf e}_- + i{\bf e}_+ \times {\bf e}_-)\cdot {\bf b} 
\rangle
C^{bj},
\label{Eq1_C0}\\
%\end{eqnarray}
%
%\begin{eqnarray}
&&
C^{aj} = \frac{\tau^2}{\tau_0}\delta_{aj} 
+ 
\frac{\tau}{2\tau_0} 
\langle 
({\bf e}_+ - {\bf e}_- - i{\bf e}_+ \times {\bf e}_-)\cdot {\bf a} 
\rangle
C^{0j}
\nonumber\\
&&
+ 
\frac{\tau}{2\tau_0}
\sum\limits_{b=x,y,z}
\langle
(1 + {\bf e}_+\cdot{\bf e}_-)({\bf a\cdot b})
-i({\bf e}_+ + {\bf e}_-)\cdot{\bf a\times b}
\nonumber\\
&&
-({\bf e}_+ \cdot {\bf a})({\bf e}_- \cdot {\bf b} ) 
-({\bf e}_- \cdot {\bf a})({\bf e}_+ \cdot {\bf b} ) 
\rangle
C^{bj}.
\label{Eq1_Ca}
\end{eqnarray}
We separated the singlet $C^{0j}$ and triplet $C^{(a,b) j}$ Cooperons with respect to the first index so that $a,b$ run over $x,y,z$ only. 
Respectively, vectors ${\bf a}$ and ${\bf b}$ run over the unit vector basis of the Cartesian system. 
We also introduced a convenient shorthand notation
\begin{eqnarray}
{\bf e}_\pm = {\bf e}_{ \pm k_F\cdot{\bf n} }.
\label{e_pm}
\end{eqnarray}
As discussed in subsection~\ref{ss_H_HgTe}, the specifics of the effective Hamiltonian for HgTe quantum wells consists in the symmetry of the mass term (\ref{M_HgTe}). 
Being an even function of ${\bf k }$, it breaks the symmetry of Hamiltonian $H$ in Eq.~(\ref{H}) 
under transformation ${\bf k},\mbox{\boldmath$\sigma$} \to -{\bf k},-\mbox{\boldmath$\sigma$}$. 
The symmetry breaking is encoded in the unit vectors ${\bf e}_\pm$ (\ref{e_pm}) which have {\em antiparallel} in-plane and {\em parallel} out-of-plane components, 
\begin{eqnarray}
{\bf e}_\pm=\pm {\bf e}_{_\|} + {\bf e}_\perp, 
\label{e_HgTe}
\end{eqnarray}
where ${\bf e}_{_\|}$ and ${\bf e}_\perp$ are defined by Eq.~(\ref{e_pp}).  
In view of the identities
\begin{eqnarray}
&&
{\bf e}_+ + {\bf e}_-=2{\bf e}_\perp,
\quad
{\bf e}_+ - {\bf e}_-=2{\bf e}_{_\|}, 
\nonumber\\
&&
{\bf e}_+ \cdot {\bf e}_-=1-2{\bf e}^2_{_\|}, 
\quad
{\bf e}_+ \times {\bf e}_-=2{\bf e}_{_\|}\times{\bf e}_\perp, 
\label{ee_HgTe}
\end{eqnarray}
Eqs. (\ref{Eq1_C0}) and (\ref{Eq1_Ca}) reduce to 
\begin{eqnarray}
&&
C^{0j} = 
\frac{\tau^2}{\tau_0}\delta_{0j} 
+ 
\frac{\tau}{\tau_0} 
\langle 
{\bf e}^2_{_\|}    
\rangle
C^{0j}
\label{Eq_C0_HgTe}\\
&&
+ 
\frac{\tau}{\tau_0}
\sum\limits_{b=x,y,z} 
\langle 
({\bf e}_{_\|} + i{\bf e}_{_\|} \times {\bf e}_\perp)\cdot {\bf b} 
\rangle
C^{bj},
\nonumber\\
%\end{eqnarray}
%
%\begin{eqnarray}
&&
C^{aj} = \frac{\tau^2}{\tau_0}\delta_{aj} 
+ 
\frac{\tau}{\tau_0} 
\langle 
({\bf e}_{_\|} - i{\bf e}_{_\|} \times {\bf e}_\perp)\cdot {\bf a}  
\rangle
C^{0j}
\label{Eq_Ca_HgTe}\\
&&
+ 
\frac{\tau}{\tau_0}
\sum_{b=x,y,z}
\langle
{\bf e}^2_\perp ({\bf a\cdot b})
-i{\bf e}_\perp\cdot ({\bf a\times b})
+({\bf e}_{_\|} \cdot {\bf a})({\bf e}_{_\|} \cdot {\bf b} ) 
\nonumber\\
&&
-
({\bf e}_\perp \cdot {\bf a})({\bf e}_\perp \cdot {\bf b} ) 
\rangle
C^{bj},
\nonumber
\end{eqnarray}
or, explicitly,
\begin{eqnarray}
\left[\frac{\tau_0}{\tau} - \langle e^2_{_\|}  \rangle  \right]C^{0j}
&-& 
\langle e_x + ie_ye_\perp \rangle C^{xj}
\label{C0}\\
&-& 
\langle e_y - ie_xe_\perp \rangle C^{yj}
=\tau\delta_{0j}, 
\nonumber
\end{eqnarray}
\begin{eqnarray}
-\langle  e_x - ie_ye_\perp \rangle C^{0j}
&+& 
\left[\frac{\tau_0}{\tau} - \langle e^2_\perp + e^2_x \rangle \right]C^{xj}
\label{Cx}\\
&-& 
\langle e_xe_y - ie_\perp \rangle C^{yj}
=\tau\delta_{xj}, 
\nonumber
\end{eqnarray}
\begin{eqnarray}
-\langle  e_y + ie_xe_\perp \rangle C^{0j}
&-& 
\langle e_ye_x + ie_\perp \rangle C^{xj}
\label{Cy}\\
&+& 
\left[\frac{\tau_0}{\tau} - \langle e^2_\perp + e^2_y\rangle \right] C^{yj}
=\tau\delta_{yj}, 
\nonumber
\end{eqnarray}
\begin{eqnarray}
C^{zj}=\frac{\tau^2}{\tau_0}\delta_{zj}.
\label{Cz}
\end{eqnarray}
For the quantum-interference conductivity corrections we will only need the ${\bf q}$- and $\omega$-dependent diagonal Cooperons $C^{00}$, $C^{xx}$ and $C^{yy}$. 
Each of them is obtained from coupled Eqs.~(\ref{C0}) -- (\ref{Cy}) where index $j$ should be set to $0, x$ or $y$, respectively. 
The coefficients in these equations are evaluated by expanding~\cite{Altshuler80} the denominator in Eq.~(\ref{Angle}) 
in the small Cooperon momentum ${\bf q}$ and frequency $\omega$,
\begin{equation}
\tau v_{_F}{\bf n}\cdot{\bf q} \ll 1, \qquad \tau\omega \ll 1. 
\label{small}
\end{equation}
In doing so, we keep the lowest order terms that yield the nonzero angle average $\langle...\rangle$ 
and compete with the small symmetry-breaking parameter ${\bf e}_\perp$ [see, Eq.~(\ref{e_pp})]. 
Under these approximations we obtain the following expressions for the diagonal Cooperons:
\begin{eqnarray}
C^{00}({\bf q},\omega)=
\frac{1}{  D {\bf q}^2 + \tau^{-1}_{\cal M} - i\omega  }, \quad 
\tau^{-1}_{\cal M}=\frac{2 {\bf e}^2_\perp }{\tau}, 
\label{C_00}
\end{eqnarray}
\begin{eqnarray}
D=D_\tau
(2 + 5 {\bf e}^2_\perp + {\bf e}^4_\perp)/{\bf e}^2_{_\|}, \qquad\quad 
D_\tau= v^2_{_F} \tau/2,
\label{D}
\end{eqnarray}
\begin{eqnarray}
&&
C^{xx}({\bf q},\omega)=
\frac{2\tau}{ {\bf e}^2_{_\|} } 
\frac{ 2 - {\bf e}^4_{_\|}\cos^2\phi_{\bf q} }{2 + 5 {\bf e}^2_\perp + {\bf e}^4_\perp}
\label{C_xx}\\
&&
+
\frac{ 2\tau (\tau^{-1}_{\cal M} -i\omega) }{ {\bf e}^4_{_\|} } 
\frac{ (1+3{\bf e}^2_\perp)^2 - {\bf e}^6_{_\|}\sin^2\phi_{\bf q} }{2 + 5 {\bf e}^2_\perp + {\bf e}^4_\perp}
\, C^{00}({\bf q},\omega),
\nonumber
\end{eqnarray}

\begin{eqnarray}
&&
C^{yy}({\bf q},\omega)=
\frac{2\tau}{ {\bf e}^2_{_\|} } 
\frac{ 2 - {\bf e}^4_{_\|}\sin^2\phi_{\bf q} }{2 + 5 {\bf e}^2_\perp + {\bf e}^4_\perp}
\label{C_yy}\\
&&
+
\frac{ 2\tau (\tau^{-1}_{\cal M} -i\omega) }{ {\bf e}^4_{_\|} } 
\frac{ (1+3{\bf e}^2_\perp)^2 - {\bf e}^6_{_\|}\cos^2\phi_{\bf q} }{2 + 5 {\bf e}^2_\perp + {\bf e}^4_\perp}
\, C^{00}({\bf q},\omega),
\nonumber
\end{eqnarray}
where angle $\phi_{\bf q}$ in Eqs. (\ref{C_xx}) and (\ref{C_yy}) indicates the Cooperon momentum direction: 
${\bf q}=|{\bf q}| \cdot (\cos\phi_{\bf q}, \sin\phi_{\bf q}, 0)$. 

Let us analyze Eqs.~(\ref{C_00}) -- (\ref{C_yy}).
The symmetry breaking has a three-fold effect on the Cooperons. 
First, it results in a relaxation gap $\tau^{-1}_{\cal M}$ in the singlet Cooperon $C^{00}$ (\ref{C_00}), 
which implies suppression of the quantum interference for times larger than $\tau_{\cal M}$ 
(even in the absence of the phase breaking, i.e. for $\omega\to 0$).    
Second, the symmetry breaking affects the diffusion constant $D$ in Eq.~(\ref{D}). 
The diffusion constant renormalization comes from the off-diagonal Cooperons.~\cite{McCann06} 
In the absence of the symmetry breaking (i.e. for ${\bf e}^2_\perp=0$) one finds~\cite{McCann06,Ando05} 
$D= 2 D_\tau =v^2_{_F}\tau_{tr}/2$ with $\tau_{tr}=2\tau$.
Finally, the expressions for the triplet Cooperons $C^{xx}$ (\ref{C_xx}) and $C^{yy}$ (\ref{C_yy})
contain additional terms $\propto \tau/\tau_{\cal M}=2{\bf e}^2_\perp$, remaining finite in the limit $\omega\to 0$. 
Despite the smallness of the parameter ${\bf e}^2_\perp$, these terms give a noticeable contribution to the net conductivity correction. 
We will return to this point when discussing Eq.~(\ref{alpha}) in the next subsection.  

\subsection{Hikami boxes and net conductivity correction}
\label{ss_ds_HgTe}

We now turn to the evaluation of the Hikami boxes for the conductivity corrections, shown in Fig.~\ref{Diagrams}b. 
With the Cooperon defined by Eq.~(\ref{Eq}) the first and second diagrams in Fig.~\ref{Diagrams}b 
correspond to the following analytical expressions: 

\begin{widetext}
\begin{eqnarray}
\delta\sigma^{(1)}_{xx}=
\frac{e^2\hbar}{\pi N \tau^2}
\int\frac{d\epsilon}{2\pi\omega} [ f(\epsilon) - f(\epsilon+\hbar\omega) ]
&& \int\frac{ d{\bf q} }{(2\pi)^2} C_{\beta\beta^\prime\gamma\gamma^\prime}({\bf q},\omega) 
\nonumber\\
&&\times
\int\frac{ d{\bf k} }{(2\pi)^2}
(
G^{^A}_{{\bf k}, \epsilon}\,
{\cal V}^x_{\bf k}\,
G^{^R}_{ {\bf k}, \epsilon+\hbar\omega }
)_{\gamma^\prime\beta}
(
G^{^R}_{ {\bf q-k}, \epsilon+\hbar\omega }\,
{\cal V  }^x_{\bf q-k}\,
G^{^A}_{ {\bf q-k}, \epsilon }
)_{\gamma\beta^\prime},
\label{dS_1}
\end{eqnarray}
\begin{eqnarray}
\delta\sigma^{(2)}_{xx} =
\frac{e^2\hbar^2}{\pi^2 N^2 \tau_0\tau^2}
&&
\int\frac{d\epsilon}{2\pi\omega} [ f(\epsilon) - f(\epsilon+\hbar\omega) ]
\int\frac{ d{\bf q} }{(2\pi)^2} C_{\beta\beta^\prime\gamma\gamma^\prime}({\bf q},\omega)
\nonumber\\
&&\times
\int\frac{ d{\bf k} }{(2\pi)^2} \int\frac{ d{\bf k}^\prime }{(2\pi)^2}
(
G^{^A}_{{\bf k}, \epsilon}\,
{\cal V}^x_{\bf k}\,
G^{^R}_{ {\bf k}, \epsilon+\hbar\omega }
G^{^R}_{ {\bf q - k^\prime}, \epsilon+\hbar\omega }
)_{\gamma^\prime\beta}
(
G^{^R}_{ {\bf q-k}, \epsilon+\hbar\omega }\,
G^{^R}_{ {\bf k^\prime}, \epsilon+\hbar\omega }\,
{\cal V  }^x_{\bf k^\prime}\,
G^{^A}_{ {\bf k^\prime}, \epsilon }
)_{\gamma\beta^\prime},
\label{dS_2}
\end{eqnarray}
\end{widetext}
where ${\cal V}^x_{\bf k}$ is the matrix current vertex renormalized by disorder (\ref{Corr}) in the usual ladder approximation 
(see, e.g. Ref.~\onlinecite{Rammer} and diagram in Fig.~\ref{Diagrams}d) and $f(\epsilon)$ is the Fermi distribution function. 
We will skip the details regarding the third diagram in Fig.~\ref{Diagrams}a 
since it finally gives the same result as Eq.~(\ref{dS_2}). 

Evaluation of the ${\bf k}$ integrals in Eqs.~(\ref{dS_1}) and (\ref{dS_2}) yields (see, also, Appendix~\ref{B}): 
\begin{widetext}
\begin{eqnarray}
\delta\sigma^{(1)}_{xx} = 
\frac{e^2D_\tau}{2hv^2_{_F}}
%\times
%\nonumber\\
\overline{ 
[(\sigma^0 + \mbox{\boldmath$\sigma$}{\bf e}_{ k_F \cdot {\bf n}  } ) 
{\cal V}^x_{ k_F\cdot {\bf n}  }
(\sigma^0 + \mbox{\boldmath$\sigma$}{\bf e}_{ k_F  \cdot {\bf n}  } )
]_{\gamma^\prime\beta}
[(\sigma^0 + \mbox{\boldmath$\sigma$}{\bf e}_{-k_F \cdot {\bf n}  } ) 
{\cal V}^x_{-k_F \cdot {\bf n} }
(\sigma^0 + \mbox{\boldmath$\sigma$}{\bf e}_{ -k_F \cdot {\bf n}  } )
]
}_{\gamma\beta^\prime}
\int\frac{d{\bf q}}{(2\pi)^2}C_{\beta\beta^\prime\gamma\gamma^\prime}({\bf q}),\,\,
\label{dS_1_1}
\end{eqnarray}
\begin{eqnarray}
\delta\sigma^{(2)}_{xx}=\frac{e^2D_\tau\tau}{8hv^2_{_F}\tau_0}
&\times&
\overline{
[(\sigma^0 + \mbox{\boldmath$\sigma$}{\bf e}_{ k_F\cdot {\bf n} } ) 
{\cal V}^x_{ k_F\cdot {\bf n} }
(\sigma^0 + \mbox{\boldmath$\sigma$}{\bf e}_{ k_F\cdot {\bf n} } )
]_{\gamma^\prime\beta_1}
(\sigma^0 + \mbox{\boldmath$\sigma$}{\bf e}_{- k_F\cdot {\bf n}  } )
}_{\gamma\gamma_1}
\nonumber\\
&\times&
\overline{ 
(\sigma^0 + \mbox{\boldmath$\sigma$}{\bf e}_{- k_F\cdot {\bf n}^\prime } )_{\beta_1\beta}
[(\sigma^0 + \mbox{\boldmath$\sigma$}{\bf e}_{ k_F\cdot {\bf n}^\prime }) 
{\cal V}^x_{  k_F\cdot {\bf n}^\prime  }
(\sigma^0 + \mbox{\boldmath$\sigma$}{\bf e}_{ k_F\cdot {\bf n}^\prime } )
]
}_{\gamma_1\beta^\prime}
\int\frac{d{\bf q}}{(2\pi)^2}C_{\beta\beta^\prime\gamma\gamma^\prime}({\bf q}),
\label{dS_2_1}
\end{eqnarray}
\end{widetext}
where the bar denotes averaging over the momentum directions ${\bf n}$ on the Fermi surface: 
$\overline{(...)}=\int^{2\pi}_0...d\phi_{\bf n}/2\pi$. We note that the correction $\delta\sigma^{(2)}_{xx}$ 
is entirely due to the carrier chirality. If we omit the chirality matrix $\mbox{\boldmath$\sigma$}{\bf e}_{ k_F\cdot {\bf n} }$ in Eq.~(\ref{dS_2_1}), 
the independent averaging  of the current vertices ${\cal V}^x_{  k_F\cdot {\bf n}  }$ gives $\delta\sigma^{(2)}_{xx}=0$.  

The renormalized vertex ${\cal V}^x_{{\bf k}_F}$ acquires the standard prefactor $\tau_{tr}/\tau$, where $\tau_{tr}$ is 
the transport scattering time:
\begin{eqnarray}
\frac{1}{\tau_{tr}} &=& \frac{N}{\hbar}\int d\phi_{ {\bf n}^\prime } ( 1 - {\bf n}\cdot{\bf n}^\prime ) 
\zeta_{ k_F |{\bf n} - {\bf n}^\prime| }
\nonumber\\
&\times&
\frac{
1  + {\bf e}_{\perp {\bf n}} \cdot {\bf e}_{\perp {\bf n}^\prime} + 
     {\bf e}_{_\| {\bf n}} \cdot {\bf e}_{_\| {\bf n}^\prime} 
}{2}
\label{tau_tr}\\
&=&
\frac{1 + 3{\bf e}^2_\perp}{2\tau_0}.
\label{tau_tr_HgTe}
\end{eqnarray}
and satisfies the identity  
\begin{eqnarray}
(\sigma^0 + \mbox{\boldmath$\sigma$}{\bf e}_{ k_F\cdot {\bf n}  } ) 
{\cal V}^x_{ k_F\cdot {\bf n} }
(\sigma^0 + \mbox{\boldmath$\sigma$}{\bf e}_{ k_F\cdot {\bf n} } )
&=& 2 \frac{ \tau_{tr} }{ \tau } v_{_F} n_x
\label{vertex}\\
&\times& 
(\sigma^0 + \mbox{\boldmath$\sigma$}{\bf e}_{ k_F\cdot {\bf n} } ),
\nonumber
\end{eqnarray}
which helps to perform the averaging in Eqs.~(\ref{dS_1_1}) and (\ref{dS_2_1}). 
The conductivity corrections $\delta\sigma^{(1)}_{xx}$ and $\delta\sigma^{(2)}_{xx}$
can then be expressed in terms of the singlet and triplet Cooperons as follows
\begin{eqnarray}
\delta\sigma^{(1)}_{xx} &=&
\frac{2e^2D_\tau}{\pi\hbar}\Bigl( \frac{ \tau_{tr} }{ \tau } \Bigr)^2
\int\frac{ d{\bf q}  }{(2\pi)^2}
\bigl[
2 \overline{ n^2_x } (1-{\bf e}^2_\perp) C^{00}({\bf q})
\nonumber\\
&-&
2 \overline{ n^2_x (1-{\bf e}^2_y) }  C^{xx}({\bf q})
- 
2 \overline{ n^2_x (1-{\bf e}^2_x) }  C^{yy}({\bf q})
\bigr],\quad\,\,
\label{dS_1_HgTe}
\end{eqnarray}
\begin{eqnarray}
\delta\sigma^{(2)}_{xx} &=&
-\frac{2e^2D_\tau}{\pi\hbar} \frac{ \tau^2_{tr} }{ \tau\tau_0 }\, {\bf e}^2_{_\|} 
\, (\overline{ n^2_x })^2
\int\frac{ d{\bf q}  }{(2\pi)^2}
\nonumber\\
&\times&
\bigl[
(1 + {\bf e}^2_\perp) C^{00}({\bf q}) - C^{xx}({\bf q}) - {\bf e}^2_\perp C^{yy}({\bf q})
\bigr],\,\,\,
\label{dS_2_HgTe}
\end{eqnarray}
where $\overline{ n^2_x} =1/2$. We have also summed up the contributions of both Dirac cones of the HgTe QW spectrum, 
which yields the factor of $2$ in front of the integrals. Noticing further that on average over the directions of 
${\bf q}$ the triplet Cooperons (\ref{C_xx}) and (\ref{C_yy}) coincide, we express the net conductivity correction in the form: 
\begin{eqnarray}
&&
\delta\sigma_{xx}=\delta\sigma^{(1)}_{xx} + 2\delta\sigma^{(2)}_{xx} 
=
\frac{2e^2D_\tau}{\pi\hbar} \Bigl( \frac{ \tau_{tr} }{ \tau } \Bigr)^2 
\int\frac{ q dq  }{ 2\pi }
\label{dS_net_HgTe}\\
&&\times
\biggl[
\biggl( 1 - {\bf e}^2_\perp - \frac{ {\bf e}^2_{_\|} }{ 2 } \biggr) \overline{ C^{00}({\bf q}) } -  
\biggl( 1 + {\bf e}^2_\perp - \frac{ {\bf e}^2_{_\|} }{ 2 } \biggr) \overline{ C^{xx}({\bf q}) }
\biggr],
\nonumber
\end{eqnarray}
where $\overline{(...)}=\int^{2\pi}_0...d\phi_{\bf q}/2\pi$ with $\phi_{\bf q}$ defined in text after Eq.~(\ref{C_yy}). 
Note that the terms $\propto 1 \mp {\bf e}^2_\perp$ come from Eq.~(\ref{dS_1_HgTe}) for $\delta\sigma^{(1)}_{xx}$, 
whereas the terms $\propto {\bf e}^2_{_\|}/2$ come from  Eq.~(\ref{dS_2_HgTe}) for $\delta\sigma^{(2)}_{xx}$, multiplied by 2. 

\begin{figure}[b]
\includegraphics[width=70mm]{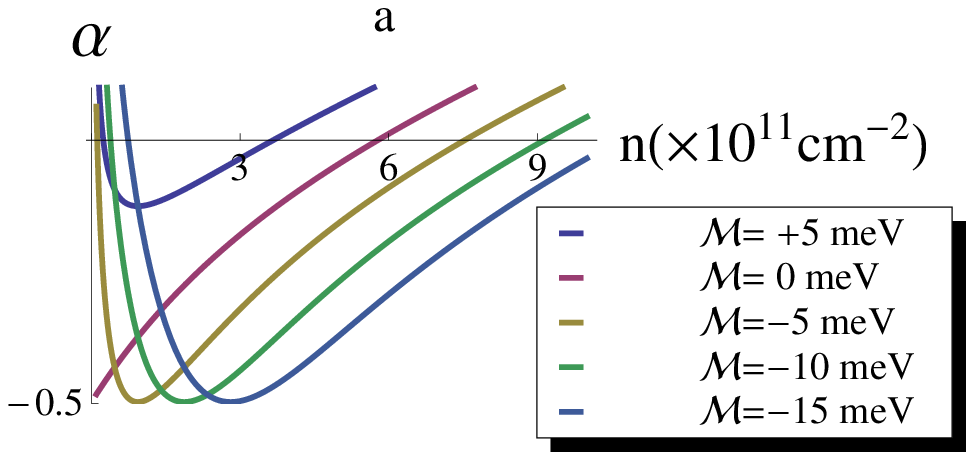}
\includegraphics[width=70mm]{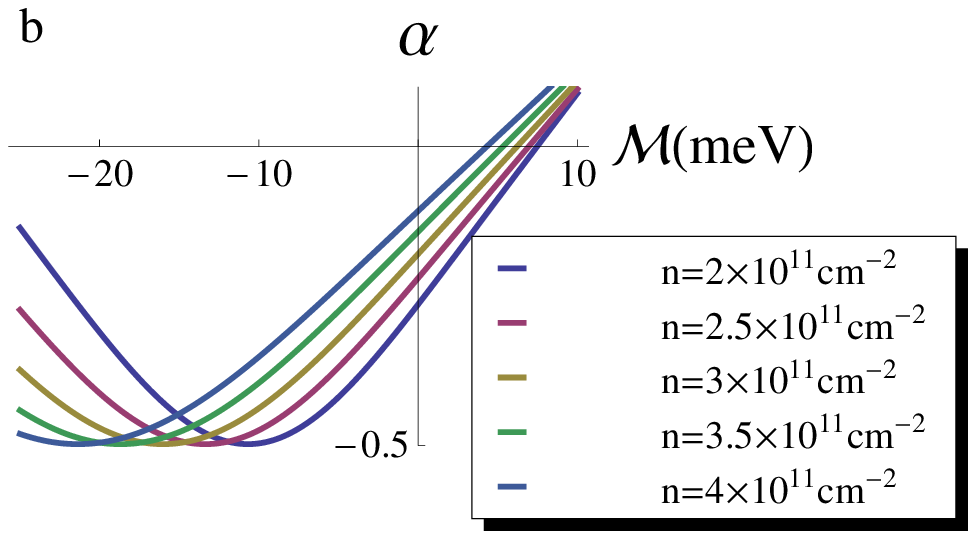}
\caption{(Color online) 
Parameter $\alpha$ (\ref{alpha}) versus carrier density $n$ (a) and band gap ${\cal M}$ (b); ${\cal A}=380$ meV$\cdot$nm, ${\cal B}=850$ meV$\cdot$nm$^2$ 
(from Ref.~\onlinecite{Buettner11}) and $\tau_0/\tau_\varphi=0.01$. 
}
\label{alpha_fig}
\end{figure}

In Eq.~(\ref{dS_net_HgTe}) the singlet Cooperon $C^{00}({\bf q})$ results in a positive conductivity correction (antilocalization) 
coming from the pairs of states with antiparallel projections of $\mbox{\boldmath$\sigma$}$, which prevents the constructive interference.~\cite{Iordanskii94,Knap96} 
This can also be viewed as the manifestation of $\pi$ Berry phase.~\cite{Suzuura02} 
In contrast, the conductivity correction due to the triplet Cooperons is negative (localization) 
because the interference of the states with the parallel projections of $\mbox{\boldmath$\sigma$}$ is constructive. 
When integrating the singlet and triplet contributions in Eq.~(\ref{dS_net_HgTe}) we insert Eq.~(\ref{C_00}) and 
the terms with the diffusion pole structure ($\propto C^{00}({\bf q})$) in Eqs.~(\ref{C_xx}) and (\ref{C_yy}).~\cite{Triplets} 
With the upper integration cut-off $(D\tau)^{-1/2}$ and replacement 
$-i\omega\to \tau^{-1}_\varphi$ in Eq.~(\ref{dS_net_HgTe}), we obtain Eq.~(\ref{dS_HgTe}) for the logarithmic correction to the Drude conductivity, 
where the relaxation gap $\tau^{-1}_{\cal M}$ is defined in Eq.~(\ref{C_00}) and the prefactor $\alpha$ is given by 
\begin{eqnarray}
&&
\alpha = 
- \frac{ {\bf e}^2_{_\|} }{ 2 + 5{\bf e}^2_\perp + {\bf e}^4_\perp } 
\times
  \left( \frac{ 1 + {\bf e}^2_\perp }{ 1 + 3{\bf e}^2_\perp } \right)^2
\label{alpha}\\
&&
\times
\left\{
{\bf e}^2_{_\|} - \tau\left( \frac{1}{ \tau_{\cal M} } + \frac{1}{\tau_\varphi} \right)
\frac{ 1 + 3{\bf e}^2_\perp }{ {\bf e}^4_{_\|} }
\frac{ 2 (1 + 3{\bf e}^2_\perp)^2 - {\bf e}^6_{_\|} }{ 2 + 5{\bf e}^2_\perp + {\bf e}^4_\perp }
\right\}.
\nonumber
\end{eqnarray}
This expression is factorized into the three parts that have different origins:
the first comes from the renormalization of the diffusion constant $D$ in Eq.~(\ref{D}), 
the second is due to the vertex renomalization [see, Eqs.~(\ref{dS_net_HgTe}), (\ref{tau_HgTe}) and (\ref{tau_tr_HgTe})], 
and the third includes the contributions of the three Hikami boxes in Fig.~\ref{Diagrams}b 
with the interplay of the singlet and triplet Cooperons [see, Eq.~(\ref{dS_net_HgTe})].
It should be noted that in the presence of the ${\bf k},\mbox{\boldmath$\sigma$} \to -{\bf k},-\mbox{\boldmath$\sigma$}$ 
symmetry there is partial cancellation of these three factors \cite{McCann06}, yielding $\alpha=-1/2$ 
for the symplectic disorder class \cite{Hikami80} (see, also Eqs.~(\ref{dS_net_BiSe0}) and (\ref{dS_net_BiSe}) for topological insulators in Sec. \ref{s_BiSe}).

Equation~(\ref{alpha}), as well as the equation for the conductivity correction (\ref{dS_HgTe}), is valid under conditions
\begin{equation}
\tau/\tau_{\cal M}=2{\bf e}^2_\perp \ll 1, \qquad  \tau/\tau_\varphi \ll 1,
\label{div}
\end{equation}
when the carrier diffusion is limited by the time-scale, ${\rm min}\{\tau_\varphi, \tau_{\cal M} \}$, longer than the elastic life-time $\tau$. 
In particular, for ${\bf e}^2_\perp = 0$ and $\tau/\tau_\varphi \to 0$ the parameter $\alpha\to -1/2$, and 
we recover the result $\delta\sigma_{xx} = 2\times \frac{e^2}{2\pi h}\ln\frac{ \tau_\varphi }{ \tau }$ 
for the symplectic class [the factor of 2 accounts for the two Dirac cones, see Eq.~(\ref{H_HgTe})]. 
For a finite ${\bf e}^2_\perp$ the broken ${\bf k},\mbox{\boldmath$\sigma$} \to -{\bf k},-\mbox{\boldmath$\sigma$}$ symmetry leads to 
the deviation of $\alpha$ from $-1/2$ [see, Fig.~\ref{alpha_fig}]. The deviation is quite significant because the expansion in powers of ${\bf e}^2_\perp$ 
involves large numerical coefficients: 
\begin{eqnarray}
\alpha \approx - \frac{1}{2} 
\left\{ 1 - \frac{ 17 {\bf e}^2_\perp}{2} - 
\frac{\tau}{2} \left( \frac{1}{ \tau_{\cal M} } + \frac{1}{\tau_\varphi} \right)
\left( 1 + \frac{35 {\bf e}^2_\perp }{2}
\right)
\right\}.
\label{alpha_exp}
\end{eqnarray}
This behavior can be seen as the crossover between the symplectic and unitary classes.

\section{Surface states in topological insulators}
\label{s_BiSe}

\subsection{Effective Hamiltonian}
\label{ss_H_BiSe}

\begin{figure}[t]
\includegraphics[width=50mm]{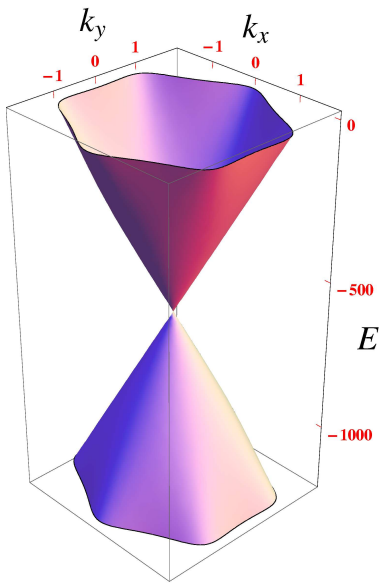}
\caption{(Color online) Energy bands of a topological surface state in meV [see, Eqs. (\ref{H_BiSe}), (\ref{M_BiSe}) and text]  
versus in-plane wave-numbers $k_x$ and $k_y$ in nm$^{-1}$. The Fermi level lies in the conduction band at $E=0$. 
We choose ${\cal A}=300$ meV$\cdot$nm, $W=50$ meV$\cdot$nm$^3$ (see, e.g., Ref.~\onlinecite{Liu10}), ${\cal D}=0$ meV$\cdot$nm$^2$, 
and $E_{_F}=600$ meV.}
\label{E_3D}
\end{figure}

\begin{figure}[t]
\includegraphics[width=50mm]{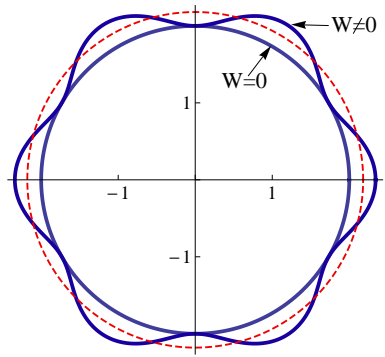}
\caption{(Color online)
Polar plots of surface-state spectrum $E(k, \phi_{\bf n})$ (in units of energy ${\cal A}k$) as a function of momentum direction
specified by angle $\phi_{\bf n}\in (0,2\pi)$ [see, also, Eq.~(\ref{Spectrum})]. 
For $W\not =0$ the spectrum shows hexagonal warping (we chose $W^2k^4/2{\cal A}^2 = 0.4$ and ${\cal D}k/{\cal A}=1$). 
On average over all angles the warping results in larger $E(k, \phi_{\bf n})$ (dashed circle) compared to the $W=0$ case (solid circle).}
\label{Warping}
\end{figure}

Unlike HgTe wells the spectrum of surface states in topological insulators (TIs), such as Bi$_2$Se$_3$ or Bi$_2$Te$_3$, 
consists of a single Dirac cone. Consequently, the effective Hamiltonian for the surface state in TIs~\cite{Fu09,Liu10} 
has the form of the single-block Hamiltonian of Eq.~(\ref{H_HgTe}):   
\begin{eqnarray}
H_{TI} = H + H_i.
\label{H_BiSe}
\end{eqnarray}
where $H$ and $H_i$ are given by Eqs.~(\ref{H}) and (\ref{H_i}), respectively.
There are two further distinctions between Hamiltonians for Bi$_2$Se(Te)$_3$ and HgTe wells. 
First, here the basis functions correspond to $\frac{1}{2}$ and $-\frac{1}{2}$ spin projections, 
i.e. Pauli ($\sigma^{x,y,z}$) and unit ($\sigma^0$) matrices act on real spin indices. 
Second, the ${\cal M}_{\bf\hat k}$-term in Hamiltonian $H$ in Eq.~(\ref{H})  
is cubic (odd) in momentum ${\bf\hat k}$:~\cite{Fu09,Liu10} 
\begin{eqnarray}
{\cal M}_{\bf\hat k }=\frac{W}{2}( \hat{k}^3_+ + \hat{k}^3_-), 
\qquad 
\hat{k}_\pm = \hat{k}_x \pm i\hat{k}_y, 
\label{M_BiSe}
\end{eqnarray}
causing no gap at ${\bf k}=0$. This term does not break the  
${\bf\hat k},\mbox{\boldmath$\sigma$} \to -{\bf\hat k},-\mbox{\boldmath$\sigma$}$ invariance,  
which is now the real time-reversal symmetry. Instead, it causes hexagonal warping of the surface-state spectrum~\cite{Fu09,Liu10} 
(see, also Figs.~\ref{E_3D} and \ref{Warping}): 
\begin{eqnarray}
E(k,\phi_{\bf n})= \sqrt{  {\cal A}^2k^2 + \frac{W^2k^6}{2}( 1 + \cos 6\phi_{\bf n})  } + {\cal D}k^2.
 \label{Spectrum}
\end{eqnarray}
We will treat the warping as weak perturbation onto the isotropic spectrum, assuming the smallness of the parameter: 
\begin{eqnarray}
\frac{W^2k^4}{2 {\cal A}^2} \ll 1.
\label{w}
\end{eqnarray}
Then the main effect of the warping consists in the increase of $E(k,\phi_{\bf n})$ on average over all angles $\phi_{\bf n}$ 
(see, dashed circle in Fig.~\ref{Warping}). This amounts to replacing ${\cal M}^2_{\bf k }$ by its angle average, 
\begin{eqnarray}
 {\cal M}^2_{\bf k } \Longrightarrow \overline{ {\cal M}^2_{\bf k } }=\frac{W^2k^6}{2},
\label{M_av}
\end{eqnarray}
in Eq.~(\ref{Spectrum}). In fact, the same replacement can be done in all {\em even} functions of 
${\cal M}_{\bf k }$, e.g. Fermi momentum, Fermi velocity, DOS etc. Then, the specific of the surface states  
is captured by the odd carrier chirality, 
$\mbox{\boldmath$\sigma$} {\bf e}_{-\bf k} = -\mbox{\boldmath$\sigma$} {\bf e}_{\bf k}$, in Eq.~(\ref{G_0}).     
Therefore, the results of the integration over the single-particle momenta ${\bf k}$ (given in Appendices~\ref{A} and \ref{B}) apply also in the present case. 
This allows us to use Eqs.~(\ref{tau}) and (\ref{tau_tr}) for the scattering times,
Cooperon equations (\ref{Eq1_C0}) and (\ref{Eq1_Ca}) as well as the Hikami boxes (\ref{dS_1_1}) and (\ref{dS_2_1}) 
to obtain the weak antilocalization conductivity corrections for the surface state in TIs.

\subsection{Disorder-averaged Green's functions and scattering times}
\label{ss_G_BiSe}

As in the case of HgTe QWs, we assume that the Fermi level lies in the conduction band of the topological surface state [Fig. \ref{E_3D}] 
and the condition of weak scattering, Eq.~(\ref{weak}), is fulfilled.  
With perturbative treatment of the warping described above, the retarded/advanced Green's functions and the elastic life-time 
for the surface state in a TI are given by Eqs.~(\ref{G_RA}) -- (\ref{e_pp}). 
The time-reversal symmetry of the surface state is encoded in the odd momentum dependence of   
the out-of-plane vector ${\bf e}_{\perp {\bf n}}$. There is no backscattering in this case~\cite{Ando05}, and 
the elastic life-time~(\ref{tau}) for isotropic $\zeta_{k_F\cdot{\bf n - n^\prime}}$ is
\begin{eqnarray}
\tau = \tau_0. 
\label{tau_BiSe}
\end{eqnarray}
In the similar way, Eq.~(\ref{tau_tr}) gives the transport scattering time as
\begin{eqnarray}
\tau_{tr}=\frac{ \tau_0 }{ 1 - {\bf e}^2_{_\|}/2 }\approx
\frac{2\tau_0}{ 1 + \overline{ {\cal M}^2_{k_F\cdot{\bf n} } }/{\cal A}^2k^2_{_F} }. 
\label{tau_tr_BiSe}
\end{eqnarray}
Its dependence on ${\bf e}_{_\|}$ (and, hence, on the variance $\overline{ {\cal M}^2_{k_F\cdot{\bf n} } }$) 
originates from the anisotropy of the Dirac-fermion scattering probability in momentum space. 

\subsection{Cooperon}
\label{ss_C_BiSe}

To write the Cooperon equations (\ref{Eq1_C0}) and (\ref{Eq1_Ca}) for the surface state in a TI 
we note that the time-reversal symmetry implies the identities: 
\begin{eqnarray}
&
{\bf e}_-=-{\bf e}_{+},
\quad
{\bf e}_+ - {\bf e}_-=2{\bf e}, 
&
\nonumber\\
& 
{\bf e}_+ \cdot {\bf e}_-=-1, 
\quad
{\bf e}_+ \times {\bf e}_-=0.
&
\label{ee_BiSe}
\end{eqnarray}
Equations (\ref{Eq1_C0}) and (\ref{Eq1_Ca}) therefore reduce to 
\begin{eqnarray}
&&
C^{0j}= \tau_0 \delta_{0j} 
+  
\langle 
1    
\rangle
C^{0j}
+  
\sum\limits_{b=x,y,z}
\langle 
{\bf e}\cdot {\bf b}
\rangle 
C^{bj},
\label{Eq_C0_BiSe}
\end{eqnarray}
\begin{eqnarray}
C^{aj}=\tau_0 \delta_{aj} 
&+&  
\langle 
{\bf e} \cdot {\bf a} 
\rangle 
C^{0j}
\nonumber\\
&+& 
\sum\limits_{b=x,y,z}
\langle
({\bf e} \cdot {\bf a})({\bf e} \cdot {\bf b} ) 
\rangle 
C^{bj},
\label{Eq_Ca_BiSe}
\end{eqnarray}
or, explicitly,
\begin{eqnarray}
\left[1 - \langle 1  \rangle  \right]
C^{0j}
- 
\langle e_x \rangle 
C^{xj}
- 
\langle e_y \rangle
C^{yj}
=\tau_0 \delta_{0j}, 
\label{C0_BiSe}
\end{eqnarray}
\begin{eqnarray}
-\langle  e_x \rangle
C^{0j}
+ 
\left[1 - \langle  e^2_x \rangle \right]
C^{xj}
=\tau_0 \delta_{xj}, 
\label{Cx_BiSe}
\end{eqnarray}
\begin{eqnarray}
-\langle  e_y \rangle
C^{0j}
+ 
\left[1 - \langle e^2_y \rangle \right]
C^{yj}
=\tau_0 \delta_{yj}, 
\label{Cy_BiSe}
\end{eqnarray}
\begin{eqnarray}
\left[ 1 - \langle {\bf e}^2_\perp \rangle \right]
C^{zj}=\tau_0 \delta_{zj}.
\label{Cz_BiSe}
\end{eqnarray}
In the above equations we use the short-hand notation ${\bf e}\equiv {\bf e}_+$ for the unit vector 
${\bf e}_{k_F\cdot{\bf n}}$ whose in- and out-of-plane components are given by Eq.~(\ref{e_pp}).
Solving Eqs.~(\ref{C0_BiSe}) -- (\ref{Cz_BiSe}) for the diagonal Cooperon coefficients, we have 
\begin{eqnarray}
&&
C^{00}({\bf q},\omega)=
\frac{1}{  D {\bf q}^2  - i\omega  },\qquad D =\frac{v^2_{_F}\tau_{tr}}{2}, 
\label{C_00_BiSe}\\
&& 
\overline{ C^{xx} }=\overline{ C^{yy} } = \tau_{tr} \biggl( 1 - \frac{ {\bf e}^2_{_\|} }{4} \biggr)
-\frac{ {\bf e}^2_{_\|} }{4} i\tau_{tr}\omega C^{00}({\bf q},\omega),\,\,\,\,
\label{C_xx_BiSe}\\
&&
C^{zz} = \frac{ \tau_0 }{ 1 - \langle {\bf e}^2_\perp \rangle  }. 
\label{C_zz_BiSe}
\end{eqnarray}
Note that the singlet Cooperon $C^{00}$ (\ref{C_00_BiSe}) remains gapless also in the presence of the warping because it 
does not break the time-reversal symmetry. The warping only modifies the diffusion constant $D$ through the transport 
scattering time (\ref{tau_tr_BiSe}). The triplet Cooperons $C^{xx}$ and $C^{yy}$ are already averaged, for presentation purposes, 
over the directions of ${\bf q}$. In the absence of the symmetry breaking 
[cf. Eqs.~(\ref{C_xx}) and (\ref{C_yy})] and for $\tau_{tr}/\tau_\varphi\ll 1$, the triplet Cooperons $C^{xx}$ and $C^{yy}$ as well as $C^{zz}$ 
can be neglected compared to $C^{00}$ in the conductivity corrections.

\subsection{Hikami boxes and net conductivity correction}
\label{ss_ds_BiSe}

Repeating the calculations described in subsection~\ref{ss_ds_HgTe}, 
we express the conductivity corrections $\delta\sigma^{(1)}_{xx}$ (\ref{dS_1_1}) 
and $\delta\sigma^{(2)}_{xx}$ (\ref{dS_2_1}) in terms of the diagonal Cooperons:
\begin{eqnarray}
\delta\sigma^{(1)}_{xx} &=&
\frac{e^2D_\tau}{\pi\hbar}\Bigl( \frac{ \tau_{tr} }{ \tau } \Bigr)^2
\int\frac{ d{\bf q}  }{(2\pi)^2}
\bigl[
2\langle n^2_x \rangle C^{00}
\nonumber\\
&-&
2\langle n^2_x e^2_x \rangle  C^{xx}
- 
2\langle n^2_x e^2_y \rangle  C^{yy}
-
2\langle n^2_x e^2_\perp \rangle  C^{zz}
\bigr],\quad\,\,
\label{dS_1_BiSe}
\end{eqnarray}
\begin{eqnarray}
\delta\sigma^{(2)}_{xx} &=&
-\frac{e^2D_\tau}{\pi\hbar} \Bigl( \frac{ \tau_{tr} }{ \tau } \Bigr)^2 \Bigl( \frac{ {\bf e}_{_\|} }{2} \Bigr)^2
\int\frac{ d{\bf q}  }{(2\pi)^2}
\bigl[
C^{00} - C^{xx}
\bigr],\quad\,\,\,\,
\label{dS_2_BiSe}
\end{eqnarray}
Keeping only the singlet Cooperon $C^{00}$, we obtain the following expression for the net correction:
\begin{eqnarray}
\delta\sigma_{xx} &=& \delta\sigma^{(1)}_{xx} + 2\delta\sigma^{(2)}_{xx}
\nonumber\\
&=&\frac{e^2D_\tau}{\pi\hbar} \Bigl( \frac{ \tau_{tr} }{ \tau } \Bigr)^2 
\Bigl( 1- \frac{ {\bf e}^2_{_\|} }{2} \Bigr)
\int\frac{ d{\bf q}  }{(2\pi)^2}
\, C^{00}({\bf q})
\label{dS_net_BiSe0}\\
&=&
\frac{ e^2 D }{\pi\hbar} 
\int\frac{ d{\bf q}  }{(2\pi)^2}
\, C^{00}({\bf q}).
\label{dS_net_BiSe}
\end{eqnarray}
After the partial cancellation in Eq.~(\ref{dS_net_BiSe0}): 
$D_\tau (\tau_{tr}/\tau)^2 ( 1- {\bf e}^2_{_\|}/2 )= D\, (\tau_{tr}/\tau) \, (\tau/\tau_{tr}) =D$,   
the prefactor in the final equation~(\ref{dS_net_BiSe}) depends only on the transport scattering time through the diffusion constant $D$ 
[see, Eq.~(\ref{C_00_BiSe}) and Ref.~\onlinecite{McCann06}]. 
Inserting Eq.~(\ref{C_00_BiSe}) into Eq.~(\ref{dS_net_BiSe}), integrating over ${\bf q}$ with the upper cut-off $(D\tau)^{-1/2}$, 
and replacing $-i\omega\to \tau^{-1}_\varphi$, we obtain Eq.~(\ref{dS_BiSe}), already discussed in Sec.~\ref{intro}.   

\begin{figure}[t]
\includegraphics[width=85mm]{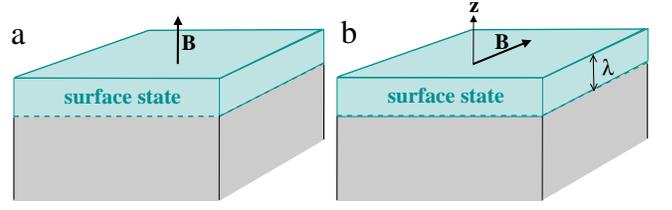}
\caption{(Color online)
Schematic view of a 3D topological insulator with a surface state subject to (a) perpendicular and (b) parallel magnetic field $B$. 
In the latter case the B-field dependence of the surface weak-antilocalization conductivity is due to the magnetic flux 
through the effective thickness of the surface state, determined by the decay length, $\lambda$, into the bulk.}
\label{Orient}
\end{figure}

Below we will focus on the dependence of $\delta\sigma_{xx}$ on the strength and oriention of an external magnetic field ${\bf B}$, 
which can be used in experiments to identify the surface states in three-dimensional topological insulators. 

\subsection{Magnetoconductivity in perpendicular field}

For ${\bf B}$ applied perpendicularly to the surface (see Fig. \ref{Orient}a) 
we write Eq.~(\ref{C_00_BiSe}) for the singlet Cooperon in the two-dimensional position representation,~\cite{Altshuler80}    
\begin{eqnarray}
\left[ D\left( i\nabla + \frac{2e}{\hbar} {\bf A}({\bf r}) \right)^2 - i\omega\right] 
C^{00}({\bf r},{\bf r}^\prime) = \delta({\bf r}-{\bf r}^\prime), 
\label{C_00_r}
\end{eqnarray}
including the vector potential ${\bf A}({\bf r})=(-By,0,0)$ of the magnetic field (${\bf r}=(x,y,0)$). 
The solution is given by the Hilbert-Schmidt expansion in the Landau wave-function basis,
which yields the well-known expression~\cite{Altshuler80} for the magnetoconductivity $\Delta\sigma_{xx}(B)=\delta\sigma_{xx}(B)-\delta\sigma_{xx}(0)$:
\begin{eqnarray}
&&
\Delta\sigma_{xx}(B)=\frac{e^2}{2\pi h}
\Biggl[
\ln\frac{ B_\perp }{B}
-\psi\Biggl(
\frac{1}{2} + \frac{ B_\perp }{B}
\Biggr)
\Biggr],
\label{dsigma_out}\\
&&
B_\perp=\frac{ \hbar }{ 2e \,\ell_{tr} \ell_\varphi },
\label{B_out}
\end{eqnarray}
where the magnetic field $B_\perp$ corresponds to the Aharonov-Bohm flux of order of $h/e$ through a typical area 
encircled by the interfering trajectories,~\cite{Altshuler80} 
$\ell_{tr}=v_{_F}\tau_{tr}$ and $\ell_\varphi=v_{_F}\tau_\varphi$, and $\psi(x)$ is the digamma function.  

\subsection{Magnetoconductivity in parallel field}

In the case of the parallel magnetic field ${\bf B}$ (see Fig. \ref{Orient}b) 
the vector potential, ${\bf A}(z)= ({\bf B} \times \hat{\bf z} ) z$, depends explicitly on the coordinate $z$ 
perpendicular to the surface ($\hat{\bf z}$ is the unit vector).    
Therefore, the penetration of the surface state into the bulk needs to be taken into account. 
To do so we first transform the diffusion operator in equation $[D{\bf q}^2 -i\omega]C^{00}({\bf q})=1$ into the {\em three-dimensional} position representation 
and then make the Peierls substitution $i \nabla_{\bf r} \to i \nabla_{\bf r} + \frac{2e}{\hbar }{\bf A}(z)$ as follows 
\begin{eqnarray}
&
a^{-1}\int d{\bf r} dz e^{ -i{\bf q}{\bf r} } \chi^*(z) 
[ D( i \nabla_{\bf r} + \frac{2e}{\hbar } {\bf A}(z))^2 - i\omega  ]
e^{ i{\bf q}{\bf r} }\chi(z)  
&
\nonumber\\
&
\times C^{00}({\bf q})=1,
&
\label{3D}
\end{eqnarray}
where the in-plane (${\bf r})$ integration goes over the surface area $a$, and the out-of-plane ($z$) integral involves 
the normalized wave function of the surface state, $\chi(z)$, which
decays exponentially for $z\to -\infty$ on the length $\lambda$ inversely proportional to the bulk band gap:~\cite{Volkov85} 
\begin{eqnarray}
 \chi(z) =\frac{e^{ z/\lambda }}{ \sqrt{ \lambda/2 } }  = \frac{e^{ -|z|/\lambda }}{ \sqrt{ \lambda/2 } } .
\label{chi}
\end{eqnarray}
In the latter form (i.e. written with $|z|$) 
this equation can formally be used in the entire space $-\infty < z < \infty$. 
This helps to simplify further calculations because the system is symmetrically extended to the other half-space, 
$0 < z < \infty$, and the $z$ integral in Eq. (\ref{3D}) can be calculated as $\int dz ... = (1/2) \int^\infty_{-\infty} dz...$. 
Having done this integration, we return to ${\bf q}$ representation of the Cooperon:
\begin{eqnarray}
&&
C^{00}({\bf q},\omega)=
\frac{1}{  D {\bf q}^2 + \tau^{-1}_B - i\omega  },
\label{C_00_B}\\
&&  
\tau^{-1}_B = 2De^2B^2\lambda^2/\hbar^2 = D \lambda^2/2\ell^4_B, 
\label{tau_B}
\end{eqnarray}
where $\tau_B$ is the time-scale for the suppression of the quantum interference by the magnetic flux 
through the thickness of the surface state, $\lambda$. The latter is assumed much smaller than the magnetic length $\ell_B=\sqrt{\hbar/2|eB|}$. 
Note that Eq. (\ref{tau_B}) has a six-time larger numerical coefficent than the result for the quasi-2D quantum wells,~\cite{Altshuler81} 
which reflects the difference in the electron confinement.  

\begin{figure}[t]
\includegraphics[width=65mm]{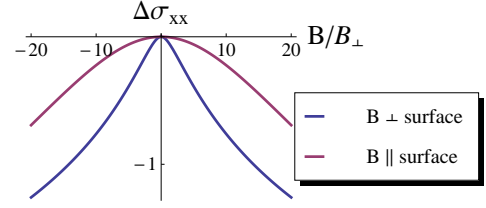}
\caption{(Color online) 
Quantum-interference magnetoconductivity $\Delta\sigma_{xx}(B)$ 
[in units of $e^2/2\pi h$, see Eqs.~(\ref{dsigma_out}) and (\ref{dsigma_in})] for in- and out-of-plane field orientations. 
The magnetic field scales, on which $\Delta\sigma_{xx}(B)$ decreases, are fixed such that  
$B_{_\|}/B_\perp = 2\sqrt{ \ell_{tr} \ell_\varphi} / \lambda = 20$.  
}
\label{dS_B_fig}
\end{figure}

Using Eqs. (\ref{dS_net_BiSe}) and (\ref{C_00_B}) we calculate the magnetoconductivity $\Delta\sigma_{xx}(B)=\delta\sigma_{xx}(B)-\delta\sigma_{xx}(0)$:
\begin{eqnarray}
\Delta\sigma_{xx}( B ) 
&=&
-\frac{e^2}{2\pi h}\ln\Biggl( 1 + \frac{\tau_\varphi}{ \tau_B } \Biggr)
\nonumber\\
&=& -\frac{e^2}{2\pi h}\ln\Biggl( 1 + \frac{B^2}{ B^2_{_\|} } \Biggr),
\label{dsigma_in}\\
B_{_\|} &=& 2 \frac{ \sqrt{\ell_{tr} \ell_\varphi} }{\lambda} \, B_\perp.
\label{B_in}
\end{eqnarray}
Clearly, for a sufficiently small penetration length $\lambda \ll 2 \sqrt{\ell_{tr} \ell_\varphi}$ the in- and out-of-plane geometries 
have distinctly different magnetic-field scales, $B_{_\|} \gg B_\perp$, on which $\Delta\sigma_{xx}(B)$ decreases with $B$ [see Fig.~\ref{dS_B_fig}]. 
The in-plane magnetoconductivity $\Delta\sigma_{xx}( B )$ (\ref{dsigma_in}) 
can be verified against recent magnetotransport measurements on Bi$_2$Te$_3$ (Ref. \onlinecite{He10}).

\begin{acknowledgments}
We thank L. W. Molenkamp, H. Buhmann, P. Brouwer, B. Trauzettel, P. Recher, C. Br\"une, C. Gould, C.-X. Liu and B. B\"uttner for helpful discussions. 
\end{acknowledgments}

\appendix
\section{Derivation of Cooperon equations (\ref{Eq4_C})}
\label{A}
The equation for the coefficients
$C^{ij}=\sum_{\alpha\beta\alpha^\prime\beta^\prime}
\Psi^{i*}_{\alpha\beta} \Psi^{j}_{\alpha^\prime\beta^\prime}\,C_{\alpha\beta\alpha^\prime\beta^\prime}$
follows from Eq.~(\ref{Eq}):
\begin{eqnarray}
&&
C^{ij}({\bf q})= 
\frac{\tau^2}{\tau_0}\delta_{ij} 
+ 
\frac{\hbar}{\pi N \tau_0} 
\sum_s\int \frac{d{\bf k}}{(2\pi)^2}
\Bigl[
\sum_{\alpha\beta\gamma^\prime\delta^\prime}
\label{Eq1_C}\\
&&
\times\Psi^{i*}_{\alpha\beta} 
G^{^R}_{\alpha\gamma^\prime}({\bf k}, \epsilon +\hbar\omega ) 
G^{^A}_{\beta\delta^\prime}({\bf q - k}, \epsilon)
\Psi^s_{\gamma^\prime\delta^\prime}
\Bigr]
C^{sj}({\bf q}).
\nonumber
\end{eqnarray}
It can be rewritten in a simpler form with the help of the identity  
%\begin{widetext}
\begin{eqnarray}
&&
\sum\limits_{\alpha\beta\gamma^\prime\delta^\prime}
\Psi^{i*}_{\alpha\beta} 
G^{^R}_{\alpha\gamma^\prime}({\bf k}, \epsilon +\hbar\omega ) 
G^{^A}_{\beta\delta^\prime}({\bf q - k}, \epsilon)
\Psi^s_{\gamma^\prime\delta^\prime}
=
\nonumber\\
&&
\sum\limits_{\delta^\prime\beta\alpha\gamma^\prime}
\left[ G^{{^A}^T}({\bf q - k}, \epsilon) \right]_{\delta^\prime\beta}
\left[ \Psi^{i\dagger} \right]_{\beta\alpha}
G^{^R}_{\alpha\gamma^\prime}({\bf k}, \epsilon +\hbar\omega ) 
\Psi^s_{\gamma^\prime\delta^\prime}
\nonumber\\
&&=
{\rm Tr}\left[
G^{{^A}^T}({\bf q - k}, \epsilon)
\Psi^{i\dagger} 
G^{^R}({\bf k}, \epsilon +\hbar\omega ) 
\Psi^s
\right], 
\label{Trace}
\end{eqnarray}
%
%\end{widetext}
%
where ${\rm Tr}$ and $T$ denote the trace and transposition operations, respectively, 
in $\mbox{\boldmath$\sigma$}$ space. We therefore have
\begin{eqnarray}
&&
C^{ij}({\bf q})= 
\frac{\tau^2}{\tau_0}\delta_{ij} 
+ 
\frac{\hbar}{\pi N \tau_0} 
\sum_s
{\rm Tr}
\Bigl[ 
\int \frac{d{\bf k}}{(2\pi)^2}  
\nonumber\\
&&
\times
G^{{^A}^T}({\bf q - k}, \epsilon)
\Psi^{i\dagger} 
G^{^R}({\bf k}, \epsilon +\hbar\omega ) 
\Psi^s
\Bigr]
C^{sj}({\bf q}).
\label{Eq2_C}
\end{eqnarray}
For the orthonormal basis functions given by 
\begin{eqnarray}
 \Psi^j=\frac{\sigma^j\sigma^y}{ \sqrt{2} }, \quad j=0,x,y,z,
\label{Psi}
\end{eqnarray}
Eq.~(\ref{Eq2_C}) assumes the following form:  
\begin{eqnarray}
&&
C^{ij}({\bf q})= 
\frac{\tau^2}{\tau_0}\delta_{ij} 
+ 
\frac{\hbar}{2\pi N \tau_0} 
\sum_s
{\rm Tr}
\Bigl[ 
\int \frac{d{\bf k}}{(2\pi)^2} 
\nonumber\\
&& 
\widetilde{G^{^A}}({\bf q - k}, \epsilon)
\sigma^i 
G^{^R}({\bf k}, \epsilon +\hbar\omega ) 
\sigma^s
\Bigr]
C^{sj}({\bf q}).
\label{Eq3_C}
\end{eqnarray}
Here the tilde denotes the operation
\begin{eqnarray}
\widetilde{G^{^A}}= \sigma^y \left[ G^{^A} \right]^T \sigma^y,
\label{Flip}
\end{eqnarray}
which flips the pseudospin: $\widetilde{ \mbox{\boldmath$\sigma$} }=\sigma^y \mbox{\boldmath$\sigma$}^T \sigma^y=-\mbox{\boldmath$\sigma$}$.
Thus, for ${\bf q}\to 0$ Eq.~(\ref{Eq3_C}) describes interference 
between the state with ${\bf k},\mbox{\boldmath$\sigma$}$ and its "time-reversed" partner with 
$-{\bf k},-\mbox{\boldmath$\sigma$}$.

Next, we evaluate the ${\bf k}$ integral in Eq.~(\ref{Eq3_C}) using the sharpness of the Green's functions at the Fermi level under condition (\ref{weak}) 
and the smallness of the Cooperon momentum ${\bf q}$ and frequency $\omega$ [see, Eq.~(\ref{small})]. 
In the denominator of $\widetilde{G^{^A}}_{\bf q-k}$ it is sufficient to keep only the linear term in 
${\bf q}$, i.e. $\xi_{\bf q-k }\approx \xi_{\bf k } -  \hbar {\bf v}{\bf q} +...$   
(which should be compared with energy $\hbar/\tau$): 
\begin{eqnarray}
\widetilde{G^{^A}}_{\bf q-k}\approx\frac{1}{2}
\frac{\sigma^0 - \mbox{\boldmath$\sigma$}{\bf e}_{\bf -k} }
{\epsilon_{_A} - \xi_{\bf k } + \hbar {\bf v}{\bf q}}. 
\label{G_A_exp}
\end{eqnarray}
At the same time, in the numerator we approximate ${\bf e}_{\bf q-k}\approx {\bf e}_{\bf -k}$, neglecting small terms 
${\cal A}q/\sqrt{ {\cal A}^2k^2_{_F} +  {\cal M}^2_{k_F} }\sim {\cal A}q/E_{_F} \ll 1$.

With Eqs.~(\ref{G_RA}), (\ref{G_A_exp}) and under condition (\ref{weak}) the ${\bf k}$ integral can be evaluated as follows 
\begin{widetext}
\begin{eqnarray}
&&
\int \frac{d{\bf k}}{(2\pi)^2} \overline{G^{^A}}({\bf q - k}, \epsilon)
\sigma^i 
G^{^R}({\bf k}, \epsilon +\hbar\omega ) 
\sigma^s=
\frac{1}{2^2}
\int\frac{ d\phi_{\bf n} }{2\pi}\int N(\xi_{\bf k}, {\bf n}) d\xi_{\bf k}
\frac{ 
(\sigma^0 - \mbox{\boldmath$\sigma$}{\bf e}_{-{\bf k} }) 
\sigma^i
(\sigma^0 + \mbox{\boldmath$\sigma$}{\bf e}_{ {\bf k} } ) 
\sigma^s
}
{
( \epsilon_{_A} - \xi_{\bf k } + \hbar {\bf v}{\bf q})
( \epsilon_{_R} - \xi_{\bf k } )
}
\label{I1}\\
&&
\approx
\frac{N}{4}
\int\frac{ d\phi_{\bf n} }{2\pi}
(\sigma^0 - \mbox{\boldmath$\sigma$}{\bf e}_{ -k_F \cdot {\bf n} } ) 
\sigma^i
(\sigma^0 + \mbox{\boldmath$\sigma$}{\bf e}_{  k_F \cdot {\bf n} } ) 
\sigma^s
\int d\xi_{\bf k}
\frac{1}
{
( \xi_{\bf k }  - \epsilon_{_A} - \hbar v_{_F} {\bf n} \cdot {\bf q} )
( \xi_{\bf k } - \epsilon_{_R} )
}
\label{I2}\\
&&
=
\frac{N}{4}
\int\frac{ d\phi_{\bf n} }{2\pi}
(\sigma^0 - \mbox{\boldmath$\sigma$}{\bf e}_{-k_F \cdot {\bf n} }) 
\sigma^i
(\sigma^0 + \mbox{\boldmath$\sigma$}{\bf e}_{ k_F \cdot {\bf n} } ) 
\sigma^s
\frac{2\pi i}
{i\hbar/\tau  + \hbar\omega - \hbar v_{_F} {\bf n} \cdot {\bf q} } 
\label{I3}\\
&&
=
\frac{2\pi N\tau}{\hbar}\times\frac{1}{4}
\int\frac{ d\phi_{\bf n} }{2\pi}
\frac{
(\sigma^0 - \mbox{\boldmath$\sigma$}{\bf e}_{-k_{_F} \cdot {\bf n} } ) 
\sigma^i
(\sigma^0 + \mbox{\boldmath$\sigma$}{\bf e}_{ k_{_F} \cdot {\bf n} } ) 
\sigma^s
}
{1 - i \tau\omega + i\tau v_{_F} {\bf n} \cdot {\bf q} },
\label{I4}
\end{eqnarray}
where the DOS $N(\xi_{\bf k}, {\bf n})$ is replaced by its Fermi surface value $N$ and, then, 
the $\xi_{\bf k}$ integral is calculated in the complex plane.  
Inserting Eq.~(\ref{I4}) into Eq.~(\ref{Eq3_C}) yields Eq.~(\ref{Eq4_C}).
\end{widetext}

\section{Evaluation of ${\bf k}$ integrals in Hikami boxes in Eqs. (\ref{dS_1}) and (\ref{dS_2})}
\label{B}

To calculate the ${\bf k}$ integral in Eq.~(\ref{dS_1}) we first expand the Green's functions 
$G^{_R}_{ {\bf q-k}, \epsilon +\hbar\omega }$ and $G^{_A}_{ {\bf q-k}, \epsilon}$ in small Cooperon momentum ${\bf q}$ 
as we did in Eq.~(\ref{G_A_exp}) of Appendix \ref{A}: 
\begin{eqnarray}
G^{^R}_{{\bf q-k},\epsilon +\hbar\omega} \approx \frac{1}{2}
\frac{\sigma^0 + \mbox{\boldmath$\sigma$}{\bf e}_{\bf -k} }
{\epsilon_{_R} - \xi_{\bf k } + \hbar\omega + \hbar {\bf v}{\bf q}}, 
\label{G_R_exp1}
\end{eqnarray}
\begin{eqnarray}
G^{^A}_{{\bf q-k},\epsilon} \approx \frac{1}{2}
\frac{\sigma^0 + \mbox{\boldmath$\sigma$}{\bf e}_{\bf -k} }
{\epsilon_{_A} - \xi_{\bf k } + \hbar {\bf v}{\bf q}}
\label{G_R_exp2}
\end{eqnarray}
Then, the integral is converted to $\int \frac{d\phi_{\bf n}}{2\pi}\int Nd\xi_{\bf k} ...$ and the $\xi_{\bf k}$ 
integration is again done in the complex plane following the same steps as in Eqs.~(\ref{I1}) -- (\ref{I4}). The final result is 
\begin{widetext}
\begin{eqnarray}
\int \frac{d{\bf k}}{(2\pi)^2}(
G^{^A}_{{\bf k}, \epsilon}\,
{\cal V}^x_{\bf k}\,
G^{^R}_{ {\bf k}, \epsilon+\hbar\omega }
)_{\gamma^\prime\beta}
&&
(
G^{^R}_{ {\bf q-k}, \epsilon+\hbar\omega }\,
{\cal V  }^x_{\bf q-k}\,
G^{^A}_{ {\bf q-k}, \epsilon }
)_{\gamma\beta^\prime}
\approx 
\frac{4\pi N\tau^3}{\hbar^3}\frac{1}{2^4}
\label{dS_1_k}\\
&&
\times
\overline{ 
[(\sigma^0 + \mbox{\boldmath$\sigma$}{\bf e}_{ k_F\cdot{\bf n} } ) 
{\cal V}^x_{ k_F\cdot{\bf n} }
(\sigma^0 + \mbox{\boldmath$\sigma$}{\bf e}_{ k_F\cdot{\bf n} } )
]_{\gamma^\prime\beta}
[(\sigma^0 + \mbox{\boldmath$\sigma$}{\bf e}_{ - k_F\cdot{\bf n} } ) 
{\cal V}^x_{- k_F\cdot{\bf n}}
(\sigma^0 + \mbox{\boldmath$\sigma$}{\bf e}_{- k_F\cdot{\bf n}} )
]
}_{\gamma\beta^\prime},
\nonumber
\end{eqnarray}
\end{widetext}
where $\overline{(...)}=\int^{2\pi}_0...d\phi_{\bf n}/2\pi $ is the averaging over the momentum direction ${\bf n}$. 
As the expression above is independent of $\epsilon$,
the energy integral in Eq.~(\ref{dS_1}) is $\hbar/2\pi$, which along with Eq.~(\ref{dS_1_k}) yields Eq.~(\ref{dS_1_1}). 

To evaluate the integrals over ${\bf k}$ and ${\bf k}^\prime$ in Eq.~(\ref{dS_2}) 
we set ${\bf q},\omega\to 0$ in the single-particle Green's functions and re-group them as follows 
\begin{widetext}
\begin{eqnarray}
\int\frac{ d{\bf k} }{(2\pi)^2} \int\frac{ d{\bf k}^\prime }{(2\pi)^2} 
&&
(
G^{^A}_{{\bf k}, \epsilon}\,
{\cal V}^x_{\bf k}\,
G^{^R}_{ {\bf k}, \epsilon+\hbar\omega }
G^{^R}_{ {\bf q - k^\prime}, \epsilon+\hbar\omega }
)_{\gamma^\prime\beta}
(
G^{^R}_{ {\bf q-k}, \epsilon+\hbar\omega }\,
G^{^R}_{ {\bf k^\prime}, \epsilon+\hbar\omega }\,
{\cal V  }^x_{\bf k^\prime}\,
G^{^A}_{ {\bf k^\prime}, \epsilon }
)_{\gamma\beta^\prime}
\approx
\label{I_2_kk}\\
&&
\approx
\int\frac{ d{\bf k} }{(2\pi)^2} 
[
G^{^A}( {\bf k} )
{\cal V}^x( {\bf k} )
G^{^R}( {\bf k} )
]_{\gamma^\prime\beta_1}
G^{^R}_{\gamma\gamma_1}(-{\bf k} )
\int\frac{ d{\bf k}^\prime }{(2\pi)^2}
G^{^R}_{\beta_1\beta}( -{\bf k^\prime} )
[
G^{^R}({\bf k^\prime})
{\cal V }^x({\bf k^\prime})
G^{^A}({\bf k^\prime})
]_{\gamma_1\beta^\prime}.
\nonumber
\end{eqnarray}
Each of the integrals can now be done in the similar way as in Eqs.~(\ref{I1}) -- (\ref{I4}) of Appendix \ref{A}: 
\begin{eqnarray}
\int\frac{ d{\bf k} }{(2\pi)^2} [
G^{^A}( {\bf k} )
{\cal V}^x( {\bf k} )
G^{^R}( {\bf k} )
]_{\gamma^\prime\beta_1}
G^{^R}_{\gamma\gamma_1}(-{\bf k} )
&\approx& -\frac{2\pi iN\tau^2}{\hbar^2}\frac{1}{2^3}
\nonumber\\
&\times&
\overline{ 
[(\sigma^0 + \mbox{\boldmath$\sigma$}{\bf e}_{ k_F\cdot{\bf n} }  ) 
{\cal V}^x_{ k_F\cdot{\bf n} }
(\sigma^0 + \mbox{\boldmath$\sigma$}{\bf e}_{ k_F\cdot{\bf n} } )
]_{\gamma^\prime\beta_1}
(\sigma^0 + \mbox{\boldmath$\sigma$}{\bf e}_{-k_F\cdot{\bf n} } )
}_{\gamma\gamma_1}
,
\label{I_2_k}\\
\int\frac{ d{\bf k}^\prime }{(2\pi)^2} G^{^R}_{\beta_1\beta}( -{\bf k^\prime} )
[
G^{^R}({\bf k^\prime})
{\cal V }^x({\bf k^\prime})
G^{^A}({\bf k^\prime})
]_{\gamma_1\beta^\prime}
&\approx& -\frac{2\pi iN\tau^2}{\hbar^2}\frac{1}{2^3}
\nonumber\\
&\times&
\overline{ 
(\sigma^0 + \mbox{\boldmath$\sigma$}{\bf e}_{- k_F\cdot{\bf n}^\prime } )_{\beta_1\beta}
[(\sigma^0 + \mbox{\boldmath$\sigma$}{\bf e}_{ k_F\cdot{\bf n}^\prime } ) 
{\cal V}^x_{ k_F\cdot{\bf n}^\prime }
(\sigma^0 + \mbox{\boldmath$\sigma$}{\bf e}_{ k_F\cdot{\bf n}^\prime } )
]
}_{\gamma_1\beta^\prime}.
\label{I_2_k^prime}
\end{eqnarray}
Inserting these into Eq.~(\ref{dS_2}) we obtain Eq.~(\ref{dS_2_1}). 
\end{widetext}


\begin{thebibliography}{99}
\bibitem{Triplets}
Under conditions (\ref{div}) the contribution of the first terms in Eqs.~(\ref{C_xx}) and (\ref{C_yy}) is smaller than the logarithmic correction (\ref{dS_HgTe}).
\end{thebibliography}
\end{document}